\documentclass[aps,prb,twocolumn,groupedaddress,showpacs,floatfix]{revtex4}

\usepackage{amssymb}
\usepackage{amsmath}
\usepackage{graphicx}
\begin{document}

\title{Free-energy distribution functions for the randomly forced 
directed polymer}

\author{V.S.\ Dotsenko$^{\, a}$, V.B.\ Geshkenbein$^{\, b,c}$, D.A.\
Gorokhov$^{\, b}$, and G.\ Blatter$^{\, b}$}

\affiliation{$^a$LPTL, Universit\'e Paris VI, 75252 Paris, France}

\affiliation{$^b$Theoretische Physik, ETH-Zurich, 8093 Zurich,
   Switzerland}

\affiliation{$^c$L.D.\ Landau Institute for Theoretical Physics,
   119334 Moscow, Russia}

\date{\today}

\begin{abstract}

We study the $1+1$-dimensional random directed polymer problem, i.e., an
elastic string $\phi(x)$ subject to a Gaussian random potential $V(\phi,x)$
and confined within a plane. We mainly concentrate on the short-scale and
finite-temperature behavior of this problem described by a short- but
finite-ranged disorder correlator $U(\phi)$ and introduce two types of
approximations amenable to exact solutions. Expanding the disorder potential
$V(\phi,x) \approx V_0(x) + f(x) \phi(x)$ at short distances, we study the
random force (or Larkin) problem with $V_0(x) = 0$ as well as the shifted
random force problem including the random offset $V_0(x)$; as such, these
models remain well defined at all scales. Alternatively, we analyze the
harmonic approximation to the correlator $U(\phi)$ in a consistent manner.
Using direct averaging as well as the replica technique, we derive the
distribution functions ${\cal P}_{L,y}(F)$ and ${\cal P}_L(F)$ of free
energies $F$ of a polymer of length $L$ for both fixed ($\phi(L) = y$) and
free boundary conditions on the displacement field $\phi(x)$ and determine the
mean displacement correlators on the distance $L$.  The inconsistencies
encountered in the analysis of the harmonic approximation to the correlator
are traced back to its non-spectral correlator; we discuss how to implement
this approximation in a proper way and present a general criterion for
physically admissible disorder correlators $U(\phi)$.

\end{abstract}


\maketitle

\section{Introduction}\label{sec:I}

\begin{figure}[h]
   \begin{center}
   \includegraphics[width=8.0cm]{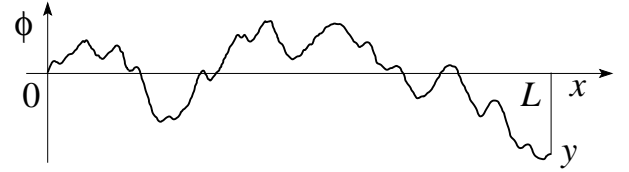}
   \caption[]{\label{fig:setup} Thermally averaged trajectory $\langle
   \phi(x)\rangle_\mathrm{th}$ of a random directed polymer in a fixed
   disorder potential $V(\phi,x)$ starting in $(x,\phi) = (0,0)$ and ending in
   $(L,y)$. The free energy associated with such a configuration is denoted by
   $F$. The random choice of the underlying disorder potential $V(\phi,x)$
   defines a random process; the free energy then turns into a random
   variable, whose distribution function ${\cal P}_{L,y}(F)$ we seek to
   calculate.}
   \end{center}
\end{figure}

Directed polymers subject to a quenched random potential have been the subject
of intense investigations during the past two decades \cite{hh_zhang_95}.
Diverse physical systems such as domain walls in magnetic films
\cite{lemerle_98}, vortices in superconductors \cite{blatter_94}, wetting
fronts on planar systems \cite{wilkinson_83}, or Burgers turbulence
\cite{burgers_74} can be mapped to this model, which exhibits numerous
non-trivial features deriving from the interplay between elasticity and
disorder.  The best understanding, so far, has been reached for the
$(1+1)$-dimensional case, i.e., a string confined to a plane, and it is this
geometry we study in the present paper. Specifically, we analyze the situation
illustrated in Fig.\ \ref{fig:setup}, an elastic string (with elasticity $c$)
of finite length $L$ within an interval $[0,L]$ directed along the $x$-axis.
The disorder potential $V(\phi,x)$ drives a finite displacement field
$\phi(x)$, which is counteracted by the elastic energy density
$c(\partial_x\phi)^2 /2$.  The problem is conveniently defined through
its Hamiltonian
\begin{equation}
   \label{1dp1}
   H[\phi(x);V] = \int_{0}^{L} dx
   \Bigl\{\frac{c}{2} \bigl[\partial_x \phi(x)\bigr]^2 
   + V[\phi(x),x]\Bigr\};
\end{equation}
the disorder potential $V(\phi,x)$ is Gaussian distributed with a zero mean
$\overline{V(\phi,x)}=0$ and a correlator
\begin{eqnarray}
   \label{1dp2}
   {\overline{V(\phi,x)V(\phi',x')}} = \delta(x-x') U(\phi-\phi'),
\end{eqnarray}
with $U(\phi)$ the correlation function. In the present work, we are mainly
interested in short-range correlated disorder potentials, which we
characterize by their extension $\xi$ and the strength $U_0 = U(0)$; these
parameters then combine in a curvature $u = -U''(0) \approx U(0)/\xi^2$. 

Quantities of central interest are the scaling behavior of the mean-squared
displacement $\overline{\langle \phi^2\rangle}(L)$ of the polymer with length
$L$ and the distribution function ${\cal P}(F)$ of the polymer's free energy
$F$. For free boundary conditions at $x= 0,L$, the former is defined through
the expression $\overline{\langle\phi^{2}\rangle}(L) \equiv \overline{\langle
[\phi(L)-\phi(0)]^2 \rangle}$, where $\langle \dots \rangle$ and
$\overline{(\dots)}$ denote thermal (temperature $T$) and disorder (random
potential $V$) averages, respectively.  Its dependence on the distance $L$
exhibits a scaling form $\overline{\langle \phi^{2} \rangle}(L) \propto
L^{2\zeta}$, with the so-called wandering exponent $\zeta$ to be
determined.\cite{numer}

The polymer's free energy $F$ is defined via its partition function
\begin{equation}
   \label{1dp5}
   Z[L,y;V] = \int_{\phi(0)=0}^{\phi(L)=y}
              {\cal D} [\phi(x)] \, \exp(-\beta H[\phi(x);V]),
\end{equation}
where $\beta = 1/T$ denotes the inverse temperature (we set the Boltzmann
constant to unity), from which the free energy
\begin{equation}
   \label{1dp6}
   F[L,y;V] = -T {\rm ln}(Z[L,y;V])
\end{equation}
follows immediately.  The free energy $F$ in Eq.\ (\ref{1dp6}) is defined for
a specific realization of the random potential $V$ and thus defines a random
variable; given the above (Gaussian) distributed disorder potential, the task
then is to determine the distribution function ${\cal P}_{L,y}(F)$.  In Eq.\
(\ref{1dp5}), we have considered a string starting at $(x, \phi) = (0,0)$ and
ending in a fixed position $(x,\phi) = (L,y)$ a distance $L$ away, but other
cases, e.g., a free boundary condition at $x=L$, see below, may be studied.

Two types of analytic solutions are known for the $(1+1)$-dimensional random
polymer problem with a short-range, in fact, $\delta$-correlated, potential
$U(\phi)=u \delta(\phi)$: mapping the replicated problem to interacting
quantum bosons \cite{kardar_87} and using the Bethe-Ansatz technique, one can
find the spectrum and eigenfunctions of the interacting quantum many-body
problem, from which the distribution function ${\cal P}_L(F)$ for the free
energy $F$ of a polymer of length $L$ and fixed endpoint at $\phi(L)=0$ can be
obtained; we call this the `longitudinal problem'. Restricting the solution to
the ground state wave function only permits the determination of the far-left
tail \cite{zhang_89}. First indications that the full distribution function
should be of the Tracy-Widom form derived from the work of Pr\"ahofer and
Spohn \cite{spohn_00} on polynuclear growth, a model in the universality class
of the Kardar-Parisi-Zhang (KPZ) model, to which the random directed polymer
problem belongs as well.  Recently, both tails of the free-energy distribution
function ${\cal P}_L(F)$ have been found using instanton techniques
\cite{Kolokolov}, with results consistent with those in Ref.\
\onlinecite{spohn_00}. Finally, the full distribution function (in the
Tracy-Widom form) has been obtained independently by several groups, by
Dotsenko\cite{Dotsenko} and by Calabrese {\it et al.} \cite{LeDoussal} using
the replica technique including the full spectrum, by Sasamoto and Spohn
\cite{Sasamoto} deriving the exact KPZ solution from the corner growth model,
and by Amir {\it et al.} \cite{Amir}.

An alternative (exact) result has been obtained using a mapping to the Burgers
equation via the Cole-Hopf transformation \cite{hhf_85}; making use of an
invariant distribution, the distribution function ${\cal P}_y (F')$ for the
free-energy difference $F'$ between two configurations with endpoints
separated by $2y$ has been found \cite{hhf_85,parisi_90} (the
so-called `transverse' problem). Both approaches have been helpful
in finding the wandering exponent\cite{hhf_85} $\zeta = 2/3$ of transverse
fluctuations of the polymer; it is generally believed that this value is
universal for short-range correlated disorder potentials, i.e., for rapidly
decaying correlator functions $U(|\phi| \to \infty) \to 0$.

Recently, some of us \cite{dotsenko_08} have studied the joint distribution
function ${\cal P}_{L,y}(\bar F,F')$  for a polymer of length $L$ which
involves two configurations of the string ending in points separated by $2y$.
An interesting result is found for the $\delta$-correlated potential, where
${\cal P}_{L,y}(\bar F,F')={P}_{L,y} (\bar F) \, {P}_{y}(F')$ separates for
large $L$ and large negative values of the mean free energy $\bar F$, with the
factor ${P}_{L,y}(\bar F)$ reproducing Zhang's tail \cite{zhang_89} for ${\cal
P}_L(F)$ and the transverse part ${P}_y(F')$ coinciding with the stationary
distribution function ${\cal P}_y(F')$ of the Burgers problem \cite{hhf_85},
thus placing previously known results into a common context.

The above solutions focus on the $\delta$-correlated disorder potential which
exhibits a singular zero-temperature limit; the problem with a finite-width
correlator $U(\phi)$ of the random potential has remained unsolved so far.
In the present paper, we study another problem, originally proposed by Larkin
\cite{Larkin_70}, that is amenable to a complete exact solution. This problem
deals with the polymer's behavior at short distances and also exhibits a
regular low-temperature limit. The basic idea is to linearize the problem,
either by an expansion of the original random potential\cite{Larkin_70}
$V(\phi,x)$ for small values of $\phi$,
\begin{equation}
   \label{Vfphi}
   V(\phi,x) \approx V_0(x) + f(x)\phi(x),
\end{equation}
or by an expansion of the correlator $U(\phi)$, see below.  Studying the
destruction of long-range order due to the presence of quenched disorder,
i.e., the behavior of the displacement field $\phi(x)$ at large distances $x$,
the random shift $V_0(x)$ can be dropped and one arrives at the {\it Larkin}
or {\it random-force model} described by the Hamiltonian (\ref{1dp1}) with
\begin{equation}
   \label{1dp15}
   V(\phi,x) = f(x) \phi(x),
\end{equation}
where $f(x)$ denotes a (Gaussian) random force field with zero mean
$\overline{f(x)} = 0$ and a correlator
\begin{equation}
   \label{1dp14}
   \overline{f(x)f(x')} = u \, \delta(x-x').
\end{equation}
Its free-energy distribution function ${\cal P}_{y,L}(F)$ has been calculated
by Gorokhov and Blatter \cite{gorokhov-blatter} for fixed boundary conditions;
here, we extend this analysis to describe the case of free boundary
conditions.  Furthermore, in our analysis of the Larkin model as a short-scale
approximation for the random potential problem, we are not allowed to ignore
the random shift $V_0(x)$; in our study, we assume the latter to be Gaussian
correlated,
\begin{equation}
   \label{1dp14_V}
   \overline{V_0(x)V_0(x')} = U_0 \, \delta(x-x'),
\end{equation}
and uncorrelated with the force, $\overline{V_0(x)f(x')}=0$; we call this
approximation the {\it shifted random force model}. Both models not only serve
as approximations to the random polymer problem on short scales but also
describe physical problems where the underlying randomness is described by a
(shifted) random force field on all length scales. Therefore, below we will
quote results for these models for arbitrary lengths $L$.

Another approximation discussed in the literature
\cite{hh_zhang_95,parisi_90b} deals with the expansion of the correlator
$U(\phi-\phi')$ rather than the potential,
\begin{equation}
   \label{1dp8}
   U(\phi-\phi') \simeq U_{p}(\phi-\phi')
   = U_{0} - \frac{1}{2} u (\phi-\phi')^{2};
\end{equation}
also, models have been studied with {\it ad hoc} power-law expressions for the
correlator \cite{hh_zhang_95}.  Here we point out, that expanding the
correlator $U(\phi)$ of the random potential or choosing arbitrary forms for
the correlator is a problematic step, as this action may generate a
non-spectral correlator introducing an ill-defined Gaussian measure and thus
lead to an unphysical model at the very start, cf.\ Sec.\ \ref{sec:RPPC} below
for a detailed discussion.  As an immediate consequence, the mapping to the
quantum boson problem requiring integration over the random potential
$V(\phi,x)$ fails.  On the other hand, performing the integration over the
random potential $V(\phi,x)$ with a well defined correlator $U(\phi)$
first and expanding the resulting interacting $-\beta^2 U(\phi)$ between
bosons only thereafter is a perfectly admissible program producing {\it
identical results}; in this case, however, we know that the (perfectly well
defined) quantum boson problem {\it does not} describe the random polymer
problem on scales where the approximate quadratic correlator deviates strongly
from the original correlator.

In the following, we will discuss the {\it harmonic-correlator approximation}
of the original random polymer problem with the understanding, that the
harmonic approximation is done after the mapping to bosons. This approximation
then relates to the above shifted random force model with a force correlator
${\overline{\partial_\phi V(\phi,x)\,\partial_{\phi'} V(\phi',x')}}|_{\phi,
\phi'=0} = [-U''(0)] \, \delta(x-x') = u\,\delta(x-x')$ and the shift
described by $\overline{V(\phi,x) V(\phi',x')}_{\phi,\phi'=0} =
U_0\,\delta(x-x')$ with the additional advantage to preserve the translation
invariance of the problem (note, that the shifted random force model involves
only the constant and mixed ($u \phi \phi'$) terms in the correlator, with the
quadratic terms $\propto(\phi^2+{\phi'}^2)$ absent). Furthermore, in Sec.\
\ref{sec:RPPC}, we will present a criterion assuring the consistent definition
of a correlator: such correlators have to be spectral.  Also, we emphasize the
difference in terminology introduced above: while the random force and shifted
random force approximations to the random potential $V(\phi,x)$ define proper
models of disordered elastic systems, this is not the case when expanding the
potential correlator $U(\phi)$; this is why we refrain from considering a {\it
model} with a harmonically correlated random potential but prefer to talk
about a harmonic {\it approximation} to the correlator.

The (shifted) random force models and the harmonic-correlator approximation
produce similar results for various quantities defined at short scales, such
as the mean free energy and the displacement correlator. While all
displacement correlators are identical for the two models and the
approximation, the random force model supplies us with the distribution
function for the {\it relaxational} free energy (i.e., the free energy of the
distorted string reduced by the energy of the straight string), whereas the
shifted random force model and the harmonic-correlator approximation provide
the distribution function for the total free energy.  Comparing the latter
two, we note that the results for the harmonic-correlator approximation
provide a more consistent description of the original problem Eq.\
(\ref{1dp2}) at short distances, as the correlator remains translation
invariant, while some terms in the expansion of the random potential are
dropped for the shifted random force problem. On the other hand, the
applicability of the harmonic-correlator approximation is limited to short
scales, as the free-energy distribution function ${\cal P}_L(F)$ suffers from
a negative second moment whenever large displacements show up, e.g., at large
distances $L$ or at high temperatures $T$; this is to be expected as the
harmonic correlator deviates strongly from the original correlator $U(\phi)$
(and eventually turns negative) at large arguments.

For a system with quenched disorder it is usually extremely difficult to find
averaged physical quantities, e.g., the mean free energy $F =-T \overline{\ln
Z}$. Replica theory, requiring calculation of the disorder-averaged $n$-th
power of the partition function $\overline{Z^n}$ then comes in as a helpful
technique. Usually, it is the limit $\lim_{n\to 0}[(\overline{Z^n}-1)/n] =
\overline{\ln Z}$, to be calculated after analytic continuation of $n$, which
is of fundamental interest. It turns out (see below) that the same quantity
$\overline{Z^n}$ and its analytic continuation is relevant in the calculation
of the free-energy distribution function ${\cal P}(F)$, since the latter is
nothing but the inverse Laplace transform of the former \cite{zhang_89}, hence
replica theory seems the technique of choice for the solution of the present
problem as well.  However, the shifted random force model defines a quadratic
problem that can be analyzed in a straightforward manner, i.e., the partition
function $Z[f(x)]$ (involving an integration over the field) can be found for
any configuration $f(x)$ of the random force and the disorder average of its
$n$-th power can be done in the end. This is opposite to the replica approach
where the integrations are interchanged, with the first integration over the
disorder of the replicated system generating an interacting imaginary-time
quantum boson problem, which then is solved in a second step (corresponding to
the integration over the field).  Below, we will discuss both procedures for
the Larkin model and find that they provide similar challenges.

The disorder ($u$ and $U_0$) and elastic ($c$) parameters of the above random
polymer problems define convenient and physically relevant length and energy
scales: The ratio of $U_0$ and $u$ defines the transverse length scale $\xi$
where the shifted random force model approximates well the random polymer
problem,
\begin{equation}
   \label{xi}
   \xi = \Bigl(\frac{U_0}{u}\Bigr)^{1/2}.
\end{equation}
Comparing the elastic energy $E_c = c \xi^2/L = c U_0/ u L $ with the disorder
energy $E_f = \sqrt{U_0 L}$ accumulated over a distance $L$, one obtains the
corresponding longitudinal scale $L_c$,
\begin{equation}
   \label{Lc}
   L_c = \Bigl(\frac{c^2 U_0}{u^2}\Bigr)^{1/3} 
   = \Bigl(\frac{c^2 \xi^2}{u}\Bigr)^{1/3}.
\end{equation}
Finally, the energy scale associated with these length scales is
\begin{equation}
   \label{Uc}
   U_c = \Bigl(\frac{c U_0^2}{u}\Bigr)^{1/3}
       = \frac{c \xi^2}{L_c}.
\end{equation}
Note that the longitudinal ($L_c$) and transverse ($\xi$) scales define the
limits of validity where our expansions describe the original random polymer
problem. The parameters are not fully appropriate to describe the results of
the Larkin model, as the latter is characterized by one disorder parameter
($u$) only---to allow for proper comparison, below, we will nevertheless
express all physical results through $\xi$, $L_c$, and $U_c$.  For the Larkin
model, these parameters will combine to expressions containing only $u$ and
$c$.

Besides providing new results for the Larkin model and a discussion of its use
as an approximation to the random potential problem at short scales, the
present study also has its merits from a methodological point of view, since
this is the only case where the entire analysis (direct and via replica) could
be carried through in a complete and consistent manner.  Below, we introduce
the formalism (Sec.\ \ref{sec:RDFB}) and then apply it to the (shifted) random
force models (Sec.\ \ref{sec:RFM}).  We then analyze the harmonic-correlator
approximation (section \ref{sec:RPPC}), analyze its failure due to its
non-spectral property, and state the spectral condition to be satisfied by a
properly defined random-potential correlator $U(\phi)$; furthermore, we
briefly present the results for the displacement correlators which are
idential in all three cases. Conclusions are presented in Sec.\ \ref{sec:D}.

\section{Methodology}
\label{sec:RDFB}

Evaluating the partition function Eq.\ (\ref{1dp5}) and the expression Eq.\
(\ref{1dp6}) for the free-energy $F$ for a given random potential $V(\phi,x)$
defines a sample-dependent random quantity. Its free-energy distribution
function ${\cal P}_{L,y}(F)$ can be derived from the $n$-th powers of
the partition function
\begin{equation}
   \label{2dp1p}
   Z^n(L,y) = \overline{Z[L,y;V]^n}
   = \overline{\exp\left(-n \beta F[L,y;V]\right)};
\end{equation}
these are equal to the (bilateral) Laplace transform of the free-energy
distribution function ${\cal P}_{L,y}(F)$ at integer multiples of
$\beta$,\cite{zhang_89}
\begin{equation}
   \label{2dp2}
   Z^n(L,y) = \int_{-\infty}^{+\infty} dF \, {\cal P}_{L,y}(F)
   \, \exp( -n \beta F).
\end{equation}
Hence, the inversion of this expression through the (inverse) Laplace
transformation provides us with the free-energy distribution function ${\cal
P}_{L,y}(F)$.  This requires us to analytically continue the expression
$Z^n(L,y)$ for the moments from integer values of $n$ to the complex
$\eta$-plane, $Z^n(L,y) \rightarrow Z(\eta;L,y)$ with $\beta n \rightarrow
\eta$. The free-energy distribution function ${\cal P}_{L,y}(F)$ then is given
by the inverse Laplace transformation
\begin{equation}
\label{2dp3}
   {\cal P}_{L,y}(F) = \frac{1}{2\pi i} \int_{-i\infty}^{+i\infty}
   d\eta \, Z(\eta;L,y) \, \exp( \eta F),
\end{equation}
where the integration goes over the imaginary $\eta$-axis with
$\mathrm{Re}(\eta)$ chosen in such a way as to place all
singularities in $Z(\eta;L,y)$ to its right. Furthermore, taking the $k$-fold
derivatives of $Z(\eta;L,y)$ with respect to $\eta$ provides us with all the
moments
\begin{equation}
   \label{2dp4}
   \overline{\langle F^{k}\rangle}(L,y) = (-1)^{k}
   \frac{\partial^k Z(\eta;L,y) }{\partial \eta^k}\Big|_{\eta=0}.
\end{equation}

The calculation of the moments $Z^n(L,y)$ involves integrations over the
displacement field $\phi(x)$ and over the distribution function $P[V(\phi,x)]$
of the disorder potential,
\begin{eqnarray}
   \label{Z_int}
   Z^n(L,y) &=& \int {\cal D} [V(\phi,x)] P[V(\phi,x)] \\
   \nonumber
   && \prod_{a=1}^n \int {\cal D} [\phi_a(x)]
   \exp\Bigl(-\beta \sum_{a=1}^n H[\phi_{a}(x)] \Bigr).
\end{eqnarray}
For the random force or Larkin model, these integrations can be done
straightforwardly in the sequence above using the distribution function
for the random force
\begin{equation}
   \label{Pf}
   P[f(x)] \propto \exp\Bigl(-\int dx\, f^2(x)/2u\Bigr);
\end{equation}
this program will be carried through in section \ref{ssec:direct} below.
Fixed and free boundary conditions are conveniently imposed by the
requirements $\phi(0)=0,~\phi(L)=y$ and $\phi(0)=0,~[\partial_x\phi](L)=0$.

On the other hand, for the general situation with a random potential
$V(\phi,x)$, the integration in Eq.\ (\ref{Z_int}) over the displacement
fields $\phi_a(x)$ cannot be done. Interchanging the integrations over
$V(\phi,x)$ and $\phi_a(x)$ takes us directly to the replica technique:
performing first the integration over the disorder potential $V(\phi,x)$,
the remaining integrations over the fields $\phi_a(x)$ have to be done
with the replica Hamiltonian $H_n[\{\phi_a\}]$,
\begin{eqnarray}
   \label{2dp5}
   \Psi(\{y_a\};L) &=& \biggl[\prod_{a=1}^{n} 
   \int_{\phi_{a}(0)=0}^{\phi_{a}(L)=y_a}
   {\cal D} [\phi_{a}(x)] \biggr]\\
   \nonumber
   &&\qquad\qquad \times
   \exp\bigl(-\beta H_{n}[\{\phi_{a}(x)\}] \bigr),\\
   \label{2dp6}
   H_{n}[\{\phi_{a}(x)\}] &=&
   \int_{0}^{L}\!\! dx \biggl\{\frac{c}{2}
   \sum_{a=1}^{n} \bigl[\partial_x \phi_{a}(x)\bigr]^2 \\
   \nonumber &&\qquad
   -\frac{\beta}{2} \sum_{a,b=1}^{n}
   U\bigl[\phi_{a}(x) -\phi_{b}(x)\bigr] \biggr\}.
\end{eqnarray}
Here, we have allowed the individual replicas of the elastic string to end in
different locations $y_a$. The expression Eq.\ (\ref{2dp5}) is identical with
the imaginary time ($x$) propagator $\Psi(\{y_a\};x)$ of a many body
problem in a path integral setting.  Collapsing the end-points $y_a = y$, this
propagator coincides with the $n$-th moment (\ref{Z_int}) of the partition
function,
\begin{equation}
   \label{2dp11}
   \Psi(\{y_a = y\}; x=L) = Z_r(n;L,y) = Z^n(L,y),
\end{equation}
where the last equation holds, provided that the integration over the disorder
potential $V$ can be exchanged with the integration over the field $\phi_a$.
Note that it is this mapping from the polymer statistical mechanics problem to
the quantum boson problem which fails when the correlator is non-spectral,
e.g., for the (naive version of) harmonic-correlator approximation.

The equivalence to a quantum many body problem becomes more obvious when going
from the path-integral Eq.\ (\ref{2dp5}) to an operator formalism; the
evaluation of the path-integral Eq.\ (\ref{2dp5}) then is equivalent to the
solution of the imaginary-time Schr\"odinger equation
\begin{equation}
   \label{2dp8}
   -\partial_x \Psi(\{y_a\};x) = \hat{H} \Psi(\{y_a\}; x)
\end{equation}
with the Hamiltonian 
\begin{equation}
   \label{2dp9}
   \hat{H} = -\frac{1}{2\beta c}\sum_{a=1}^{n}\partial_{y_a}^2
   - \frac{\beta^2}{2}\sum_{a,b=1}^{n} U(y_a-y_b).
\end{equation}
The Hamiltonian (\ref{2dp9}) describes $n$ particles of mass $\beta c$
interacting via the attractive two-body potential $-\beta^2 U(y)$; the
propagation in (\ref{2dp8}) starts in the origin at time $x=0$,
\begin{equation}
   \label{2dp10}
   \Psi(\{y_a\}; 0) = \Pi_{a=1}^{n} \delta(y_a),
\end{equation}
and ends at different coordinates $\{y_a\}$ after propagation during the time
$x$.  To keep up the formal distinction between the two quantities, we denote
the direct physical definition of the moments by $Z^n(L,y)$ (first integration
over the field $\phi$, raising the result to the power $n$, and averaging over
disorder $V$) and denote the replica expression ($n$-fold replication followed
by averaging over disorder $V$ and integration over the fields $\phi_a$ done
in the end) by $Z_r(n;L,y)$.

Finally, we note that in the replica technique, free boundary conditions at
the endpoint are more conveniently implemented through an 
integration over $y$; the partition function for the polymer with free
boundary conditions assumes the form
\begin{equation}
   \label{2dp11fV}
   Z[L;V] = N \int_{-\infty}^{+\infty}\!\! dy \,Z[L,y;V]
\end{equation}
with $N$ a suitable normalization constant.  Taking the $n$-th power and
averaging over the disorder potential $V$ provides us with the moments
$Z^n(L)$. Following the replica procedure, after replication and integration
over $V$, one arrives at the free-boundary replicated partition function
through integration over the set $\{y_a\}$ of $n$ different end-points,
\begin{equation}
   \label{2dp11f}
   Z_r(n;L) = \biggl[\prod_{a=1}^{n} N \int_{-\infty}^{+\infty} dy_a\biggr]
   \Psi(\{y_a\}; x=L).
\end{equation}
In the next sections, we are going to apply the above general schemes for the
calculation of the free-energy distribution functions ${\cal P}_{L,y}(F)$ and
${\cal P}_{L}(F)$ for fixed and free boundary conditions, respectively, of the
(shifted) random force model Eqs.\ (\ref{1dp15})--(\ref{1dp14_V}) and of the
random directed polymer model Eq.\ (\ref{1dp1}) with the parabolic
approximation for the correlation function, Eq.\ (\ref{1dp8}), done after
averaging over the disorder.  Before doing so, we briefly discuss the results
for the free string which determines our normalization $N$.

\subsection{Free string}\label{sec:FS}

The path integrals Eqs.\ (\ref{1dp5}), (\ref{Z_int}), and (\ref{2dp5}) over
trajectories $\phi(x)$ involve an arbitrary measure of integration. Here, we
choose a particular normalization such that the partition function $Z_0(L,y)$
(or wave function $\Psi_0(y,x=L)$) of the free polymer problem with fixed
boundary conditions $\phi(0)=0$ and $\phi(L)=y$ assumes the form
\begin{eqnarray}
   \label{3dp1a}
   Z_0(L,y) &=& \Psi_0(y;L) \\
   \nonumber
   &=& \int_{0}^{y} {\cal D}[\phi(x)]
   \exp\Bigl[-\frac{\beta c}{2}\int_{0}^{L} dx  
   \bigl[\partial_x \phi(x)\bigr]^2 \Bigr] 
   \\ \nonumber
   &\equiv& \exp\Bigl(-\frac{\beta c}{2 L} y^{2} \Bigr);
\end{eqnarray}
the corresponding free energy then is given by
\begin{equation}
   \label{3dp1b}
   F_0 (L,y) = \frac{c}{2 L} y^{2}.
\end{equation}
For the partition function of the free polymer with free boundary conditions
we choose the normalization $N= (2\pi L/\beta c)^{1/2}$ and obtain
\begin{equation}
   \label{3dp2a}
   Z_0(L) = \sqrt{\frac{2\pi L}{\beta c}} 
   \int_{-\infty}^{+\infty} dy \, Z_0(L,y) = 1
\end{equation}
and the free energy $F_0(L) =0$. These results will be helpful in the
interpretation of the free-energy distribution functions for the random force
model calculated below. With this normalization, all our free energies $F$ are
measured with respect to the free thermal energy $F_0^\mathrm{fs} = T
\ln\sqrt{2\pi L T/c}$ of the free string due to its entropy.

\section{Random force model}\label{sec:RFM}

We select the simplest case, the Larkin model, for the discussion of the two
methodological approaches involving either direct integration over the field
$\phi$ and subsequent disorder average over $V$ or the route following the
replica approach.  While the first route is preferably done in Fourier space,
the replica calculation will be formulated in real space. Also, note that the
analysis for the Larkin- or random force model provides the distribution
function for the {\it relaxational} free-energy $F-E_0$ rather then the
(total) free energy $F$ of the polymer,
\begin{eqnarray}
   \label{relF}
   Z[L,y;V] &=& \exp^{-\beta F[L,y;V]}
   \\ \nonumber
   \qquad &\approx&\!\!\int_{\phi(0)=0}^{\phi(L)=y}
              \! {\cal D} [\phi(x)] \, \exp(-\beta H[\phi;V_0 + f\phi])
   \\ \nonumber
   \qquad &=&\! e^{-\beta E_0}\! \int_{\phi(0)=0}^{\phi(L)=y}
              \! {\cal D} [\phi(x)] \, \exp(-\beta H[\phi;f\phi]),
\end{eqnarray}
with $E_0 = \int dx V_0(x)$ the disorder energy of a straight string. This
latter remark is relevant in the comparison of the random force and the
harmonic models.

A further speciality of the Larkin model is the separation between the thermal
and the quenched disorder \cite{denis_gorokhov_diss}.  Indeed, splitting the
displacement field $\phi(x)$ into the Hamilonian's minimizer
$\phi_\mathrm{q}(x)$,
\begin{equation}
   \label{minimizer}
   c \partial_x^2 \phi_\mathrm{q}(x) = f(x),
\end{equation}
and fluctuations $\delta\phi(x)$ around it, $\phi(x) = \phi_\mathrm{q}(x)
+\delta\phi(x)$, we can decompose the Hamilonian into the free part
$H_0[\delta\phi(x)]$ and the energy of the minimizer $H[\phi_\mathrm{q}(x)]$,
\begin{eqnarray}
   \label{split_H}
   H_0[\delta\phi(x)] &=& \int_0^L dx \,
   \frac{c}{2}[\partial_x \delta \phi(x)]^2
   \\ \nonumber
   H[\phi_\mathrm{q}(x)] &=&
   \int_0^L dx \, \Bigl\{\frac{c}{2}[\partial_x \phi_\mathrm{q}(x)]^2
   +f(x) \phi_\mathrm{q}(x)\Bigr\}.
\end{eqnarray}
In addition, we can account for the boundary condition $\phi(L) = y$ through a
simple shift  $\phi(x) \rightarrow x\, y/L +\phi(x)$, which adds the terms
\begin{eqnarray}
   \label{ham_bound}
   H_y = \frac{c y^2}{2L} +\frac{y}{L} \int_{0}^{L} dx \, x f(x)
\end{eqnarray}
to the Hamiltonian.  The partition sum then naturally separates into thermal
and quenched-disorder averaged factors,\cite{denis_gorokhov_diss}
\begin{eqnarray}
   \label{split_Z}
   \overline{Z(L,y;f)} = Z_0(L,y) \> \overline{
   \exp\bigl\{-\beta \bigl( H[\phi_\mathrm{q}(x)]+H_y \bigr)\bigr\} },
\end{eqnarray}
where the factor $Z_0(L,y)$ is the partition function Eq.\ (\ref{3dp1a}) of
the free propagation.

\subsection{Direct integration}\label{ssec:direct}

The direct integration of the partition function Eq.\ (\ref{1dp5}) for the
random force problem $V(\phi,x) = f(x) \phi(x)$ is conveniently done within a
Fourier representation. For technical convenience we extend the problem to the
interval $[-L,L]$ and define the anti-symmetric force and displacement fields
$f(-x) \equiv -f(x>0)$ and $\phi(-x) \equiv -\phi(x>0)$. The relevant
quantities in Fourier space are the sine-transforms
\begin{eqnarray}
   \label{sine_trans}
   g_m = \int_{-L}^{L} dx \,g(x) \sin(k_m x)
\end{eqnarray}
with $k_m = m\pi/L$; the back transformation reads
\begin{eqnarray}
   \label{sine_trans_back}
   g(x) = \frac{1}{L} \sum_{m=1}^{\infty} g_m(x) \sin(k_m x).
\end{eqnarray}
The Hamiltonian (on the interval $[0,L]$) assumes the form (we make use of the
solution $\phi_{\mathrm{q}m} = -f_m/c k_m^2$ of Eq.\ (\ref{minimizer}) in
Fourier representation)
\begin{eqnarray}
   \label{ham_trans}
   H[\phi_{\mathrm{q}}] + H_y
   &=& -\sum_{m=1}^\infty \frac{f_m^2}{4 c L k_m^2} \\
   \nonumber
   &&\qquad -\frac{y}{L} \sum_{m=1}^\infty \frac{(-1)^m}{k_m} f_m
\end{eqnarray}
and the partition function reads
\begin{eqnarray}
   \label{part_fu}
   Z(L,y;f) &=& 
   \exp\Bigl(-\frac{\beta c y^2}{2L}\Bigr)
   \\ \nonumber
   && \times \prod_{m=1}^{\infty}
   \exp\Bigl(\frac{\beta f_m^2}{4cLk_m^2} + \frac{\beta y}{L k_m}
   (-1)^m f_m\Bigr).
\end{eqnarray}
The disorder average in the partition function Eqs.\ (\ref{split_Z}) or
(\ref{part_fu}) has to be taken over the distribution function for the random
force $f$, cf.\ Eq.\ (\ref{Pf}), or in Fourier space,
\begin{eqnarray}
   \label{ran_force}
   P(f_m) = \frac{1}{\sqrt{4\pi uL}} \exp(-f_m^2/4uL).
\end{eqnarray}
Taking the result (\ref{part_fu}) to the $n$-th power and integrating over the
force distribution Eq.\ (\ref{ran_force}), we obtain the intermediate result
\begin{eqnarray}
   \label{part_fu_n_int}
   && Z^n(L,y) =  
   \exp\Bigl(-\frac{\beta n \, c y^2}{2L}\Bigr)
   \\ \nonumber
   && \quad \times
   \prod_{m=1}^{\infty}
   \Bigl[ 1- \frac{s}{\pi^2 m^2} \Bigr]^{-1/2}
   \exp\Bigl(\sum_{m=1}^\infty\frac{\beta n\,cy^2\, 2s}
   {2L\,\,(\pi^2 m^2-s)}\Bigr)\\
   \noalign{\vspace{5pt}}
   \label{eq:def_s} 
   && \quad\qquad\qquad \textrm{with}\qquad
   s = \frac{\beta n u L^2}{c}.
\end{eqnarray}
Using the product and partial fraction expansion of circular functions
\cite{Abram}
\begin{eqnarray}
   \label{eq:prod_sin}
   \frac{\sin\sqrt{s}}{\sqrt{s}} &=&
   \prod_{m=1}^{\infty}
   \Bigl[ 1- \frac{s}{\pi^2 m^2}\Bigr], \\
   \label{eq:sum_tan}
   \frac{\sqrt{s}}{\tan\sqrt{s}} &=& 1+ \sum_{m=1}^\infty
   \frac{2s}{s-m^2\pi^2},
\end{eqnarray}
we obtain the final result
\begin{eqnarray}
   \label{3dp26}
   Z^n(L,y) &\equiv& Z(s;\epsilon) \\
   \nonumber
   &=& \Bigl(\frac{\sqrt{s}}{\sin\sqrt{s}}\Bigr)^{1/2}
       \exp\Bigl[-\epsilon\frac{s\sqrt{s}}{\tan\sqrt{s}} 
           \Bigr],
\end{eqnarray}
with the dimensionless displacement parameter
\begin{eqnarray}
   \label{3dp39}
   \epsilon = \frac{c^2}{2u}\frac{y^2}{L^3}.
\end{eqnarray}
With our normalization, the partition sum does not depend on temperature any
more (note that in the calculation of the free-energy distribution function,
the variable $s$ will be integrated over, cf.\ Eq.\ (\ref{3dp32})). The result
Eq.\ (\ref{3dp26}) is well defined provided that $0 < s < \pi^2$; the
singularity at $s = \pi^2$ will determine the shape of the left tail in the
free-energy distribution function, see (\ref{3dp43b}) below.

As a simple application, we can use the partition function Eq.\
(\ref{part_fu}) to find the free energy $\overline{\langle F \rangle} = -T\,
\overline{\ln Z(L,0;f_m)}$ of the string starting and ending in $\phi = 0$.
Taking the disorder average over the term $\sum_m \beta f_m^2/4 c L k_m^2$ in
the logarithm of the partition function Eq.\ (\ref{part_fu}), we obtain the
result
\begin{equation}
   \label{Free_energy}
  \overline{\langle F \rangle} = -\frac{uL^2}{2c\pi^2} \sum_{m=1}^\infty
  \frac{1}{m^2} = -\frac{U_c}{12} \Bigl(\frac{L}{L_c}\Bigr)^2,
\end{equation}
where we have used the Riemann Zeta function $\zeta(2) = \pi^2/6$ and the
definitions Eqs.\ (\ref{Lc}) and (\ref{Uc}). Alternatively, we can use the
Eqs.\ (\ref{2dp4}) and (\ref{3dp26}) and calculate $\overline{\langle F
\rangle}(L,0) = -(u L^2/c)\partial_s Z(s;0)|_{s=0}$. With $Z(s;0) \approx
1+s/12$ we then easily recover the above result. Note that the result Eq.\
(\ref{Free_energy}) measures the free energy $F$ with respect to the entropic
contribution $F_0^\mathrm{fs} = T \ln\sqrt{2\pi L T/c}$ of the free string.

The result for the free boundary condition $\partial_x\phi|_{x=L} = 0$ is
obtained by using an alternative expansion: first, we symmetrically extend the
system from the interval $[0,L]$ to the interval $[0,2L]$ with the definitions
$\phi(L+x) \equiv \phi(L-x)$, $f(L+x) \equiv f(L-x)$. Second, we expand the
analysis to the interval $[-2L,2L]$ using the same anti-symmetric extension as
above. As a result, we can expand the displacement and force fields into modes
$\sin(q_m L)$ with $q_m = (2m-1)\pi/2L$, $m=1,\dots,\infty$ and hence zero
slope at $x=L$. Following the same steps as above, we arrive at Eq.\
(\ref{part_fu_n_int}) with $y=0$ and the product corresponding to the
expansion of the cosine \cite{Abram},
\begin{equation}
   \label{eq:prod_os}
   \cos\sqrt{s} = \prod_{m=1}^{\infty}
   \Bigl[ 1- \frac{4s}{\pi^2 (2m-1)^2}\Bigr].
\end{equation}
The final result for the partition function with free boundary conditions then
reads
\begin{equation}
   \label{3dp27}
   Z(s) = \frac{1}{\sqrt{\cos\sqrt{s}}},
\end{equation}
where the regime of applicability is restricted to the domain $0<s<\pi^2/4$;
again, the singularity at $s = \pi^2/4$ determines the shape of the left tail
in the free-energy distribution function, cf.\ (\ref{3dp45a}).  The
alternative procedure of realizing the free boundary condition via integration
over the end-point coordinate $y$, cf.\ Eq.\ (\ref{2dp11fV}), provides the
identical result, although via a much more tedious calculation of
determinants.

\subsection{Replica approach}\label{ssec:replica}

The replica Hamiltonian Eq.\ (\ref{2dp6}) for the random force problem Eq.\
(\ref{1dp15}) reads
\begin{eqnarray}
   \label{3dp5}
   H_{n}[\{\phi_{a}\}] &=& \frac{1}{2} \int_{0}^{L}\!\! dx
   \Bigl\{c \sum_{a=1}^{n}\bigl[\partial_x \phi_{a}(x)\bigr]^2
   \\ \nonumber
   &&\qquad\qquad - \beta u \sum_{a,b=1}^{n} \phi_{a}(x)\phi_{b}(x)\Bigr\}
   \\ \label{3dp5a}
   &=& -\frac{1}{2} \int_{0}^{L} dx \sum_{a,b=1}^{n}
   \phi_{a}(x) \, U_{ab} \, \phi_{b}(x)
\end{eqnarray}
with the matrix 
\begin{equation}
\label{3dp6}
   U_{ab} =  c \, \delta_{ab} \, \partial^{2}_x + \beta u.
\end{equation}
Accounting for the random shift $V_0(x)$, cf.\ Eq.\ (\ref{1dp14_V}), adds an
additional term $-n^2 \beta U_0 L/2$ to the Hamiltonian (\ref{3dp5}).

The matrix $U_{ab}$ can be easily diagonalized and we find one $(n-1)$-fold
degenerate eigenvalue $\lambda_{1} = c \, \partial^{2}_x$ pertinent to the
free string with the $(n-1)$ orthonormal eigenvectors $\xi^{a}_{i}$ obeying
the constraint $\sum_{a=1}^{n} \xi^{a}_{i} =0$, $i = 1, \dots, n-1$. The
$n$-th eigenvalue $\lambda_{2} = c \, \partial^{2}_x \, + \, \beta n u$ is
non-degenerate and appertains to an inverted harmonic potential problem; the
associated eigenvector is $\xi^{a}_{n} = 1/\sqrt{n}$, $a = 1, \dots, n$.  The
coefficients $\xi^{a}_{i}$ of the $(n\times n)$ transformation matrix
$\xi^{a}_{i}$ satisfy the conditions $\sum_{a=1}^{n} \xi^{a}_{i} \xi^{a}_{j} =
\delta_{ij}$ (completeness) and $\sum_{i=1}^{n} \xi^{a}_{i} \xi^{b}_{i} =
\delta_{ab}$ (orthonormality). In terms of the new fields $\varphi_{i}(x)$ and
boundary conditions $q_i = \varphi_{i}(L)$,
\begin{equation}
\label{3dp11}
   \varphi_{i}(x) = \sum_{a=1}^{n} \, \xi^{a}_{i} \, \phi_{a}(x),
   \qquad q_i = \sum_{a=1}^{n} \, \xi^{a}_{i} \,y_a,
\end{equation}
the wave function (or propagator) Eq.\ (\ref{2dp5}) takes the form 
\begin{eqnarray}
   \label{3dp12a}
   &&\Psi(\{q_i\};L) = \prod_{i=1}^{n-1}\biggl[
   \int_{0}^{q_{i}}
   {\cal D}[\varphi_{i}(x)]\biggr] \int_{0}^{q_{n}} {\cal D}[\varphi_{n}(x)] \\
   \nonumber
   &&\hspace{1.2truecm}
   \times\exp\Bigl[-\frac{\beta c}{2}\! \int_{0}^{L} \!\!\! dx \,
   \sum_{i=1}^{n-1} [\partial_x \varphi_{i}(x)]^2\Bigr] \\
   \nonumber
   &&\hspace{1.2truecm}
   \times \exp\Bigl[-\frac{\beta c}{2}\! \int_{0}^{L} \!\!\! dx\,
   [(\partial_x\varphi_{n}(x))^2-\varphi_{n}^{2}(x)/\lambda^2]\Bigr],
\end{eqnarray}
where we have introduced the length parameter $\lambda$, cf.\
Eq.\ (\ref{eq:def_s}),
\begin{equation}
   \label{3dp20}
   \lambda^2 = \frac{c}{\beta n u} = \frac{L^2}{s}.
\end{equation}

The $(n-1)$ free propagators are given by $\Psi_0(q_i;L)$, $i = 1, \dots,
n-1$, cf.\ Eq.\ (\ref{3dp1a}). The propagator $\Psi_{\rm ih}(q_n;L)$ for the
inverted harmonic potential problem is obtained by solving the imaginary-time
Schr\"odinger equation 
\begin{equation}
   \label{3dp18}
   \partial_x \Psi_{\rm ih}(q;x)=\frac{1}{2}\biggl[\frac{1}{\beta c}
   \partial_{q}^2+\frac{\beta c}{\lambda^2}\,q^2\biggr]\,\Psi_{\rm ih}(q;x)
\end{equation}
with the initial condition $\Psi_{\rm ih}(q;x=0) = \delta(q)$. With the
Gaussian Ansatz $\Psi_{\rm ih}(q;x) = \chi(x)\,\exp[-a(x) \, q^{2}/2]$
and proper accounting of the initial condition, we find the solution
(cf.\ Ref.\ \onlinecite{Zinn})
\begin{equation}
   \label{Psi^ih}
   \Psi_{\rm ih}(q_n;L)=
   \Bigl(\frac{\sqrt{s}}{\sin\sqrt{s}}\Bigr)^{1/2}\\
   \exp\Bigl(-\frac{\beta c}{2L}
   \frac{\sqrt{s}}{\tan\sqrt{s}} {q_n}^2\Bigr).
\end{equation}

Inserting the free (Eq.\ (\ref{3dp1a})) and harmonic (Eq.\ (\ref{Psi^ih}))
factors into the full propagator Eq.\ (\ref{3dp12a}) and transforming back to
original variables, $\sum_{i=1}^{n-1} q_i^2 = \sum_{i=1}^{n} q_i^2 - q_n^2 =
\sum_{a=1}^{n}y_a^2 - (1/n)(\sum_{a=1}^n y_a)^2$, we obtain the result
\begin{equation}
   \label{Psi}
   \Psi(\{y_a\};L) \!=\! \Bigl[\!\prod_{a=1}^n \Psi_0(y_a;L)\Bigr]
   \frac{\Psi_{\rm ih}(\sum_b y_b/\sqrt{n};L)}{\Psi_0
   (\sum_b y_b/\sqrt{n};L)}.
\end{equation}
Choosing the appropriate boundary conditions $y_{a}=y$, $a=1,\dots,n$, we
obtain the replica partition function Eqs.\ (\ref{2dp5}) and (\ref{2dp11})
identical to the previous result Eq.\ (\ref{3dp26}), $Z_r(s;\epsilon)=
Z(s;\epsilon)$, $0<s<\pi^2$. 

The result for the (replica) partition function has been derived for positive
integer $n$. Since $Z(s;\epsilon)$ depends on $n$ only via the parameter
$s$, the expression Eq.\ (\ref{3dp26}) can be analytically continued to the
complex half-plane restricted by the condition $\textrm{Re}[s] < \pi^2$.

For the free boundary condition, we obtain the replica partition function via
integration of Eq.\ (\ref{Psi}) over all end-points $\{y_a\}$, cf.\ Eq.\
(\ref{2dp11f}); the integration is conveniently done in the variables $q_a$
and we make use of the normalization Eq.\ (\ref{3dp2a}) to obtain the result
identical to Eq.\ (\ref{3dp27}), $Z_r(s) = 1/\sqrt{\cos\sqrt{s}}$,
$0<s<\pi^2/4$.  Furthermore, the analytic continuation to real {\it
negative} values of the parameter $n$ provides the expression
\begin{equation}
   \label{3dp30}
   Z(s) = \frac{1}{\sqrt{\cosh\sqrt{|s|}}} \qquad (s < 0);
\end{equation}
alternatively, this result is obtained via the solution of the Schr\"odinger
equation Eq.\ (\ref{3dp18}) for negative $n$ involving a summation over the
discrete spectrum of the parabolic potential, see Appendix \ref{sec:nrn}.

\subsection{Distribution function: fixed boundary condition}\label{sec:FiBC}

We now turn to the calculation of the free-energy distribution function ${\cal
P}_{L,y} (F)$ from the partition function $Z^n(L,y)$.  Following the procedure
described in Sec.\ \ref{sec:RDFB}, specifically Eqs.\ (\ref{2dp2}) and
(\ref{2dp3}), the Laplace transform and its inverse assume the form
\begin{eqnarray}
   \label{3dp31}
   Z(s;\epsilon) &=& 
   \int_{-\infty}^{+\infty} df\, p_{\epsilon} (f)\,\exp(-sf),
   \\
   \label{3dp32}
   p_{\epsilon}(f) &=& \frac{1}{2\pi i} \int_{-i\infty}^{+i\infty} 
   ds \, Z(s;\epsilon) \, \exp( s f),
\end{eqnarray}
where
\begin{eqnarray}
\label{3dp33}
   f(F,L) \!\!&=&\!\! \frac{F}{F_f(L)},\quad F_f(L) = \frac{u}{c} L^2  
   = {U_c}\Bigl(\frac{L}{L_c}\Bigr)^2\!,\\
\label{3dp33b}
   \epsilon(y,L) \!\!&=&\!\! \frac{c^2}{2u}\frac{y^2}{L^3},
\end{eqnarray}
are the rescaled free energy of the system and the rescaled displacement
parameter; the original free-energy distribution function ${\cal P}_{L,y} (F)$
then derives from the rescaled expression $p_{\epsilon}(f)$ through the
relation
\begin{equation}
   \label{3dp32PF}
   {\cal P}_{L,y} (F) = \frac{1}{F_f(L)}p_{\epsilon(y,L)}(f(F,L)).
\end{equation}
Note, that the parameter $\xi$ drops out in the combination $U_c/L_c^2$, as
has to be the case for the random force model where the disorder is
characterized by only one parameter, its strength $u$. Or in other words,
using the results below for a random force (rather then a random potential)
problem, these are valid for all length scales.  For the relaxational
free-energy distribution function of the system with fixed boundary condition,
we obtain the expression
\begin{eqnarray}
\label{3dp34}
   p_{\epsilon}(f) &=& \frac{1}{2\pi i} 
   \int_{-i\infty}^{+i\infty}\!\!\!  ds
   \, \Bigl(\frac{\sqrt{s}}{\sin\sqrt{s}}\Bigr)^{1/2}\\ \nonumber
   &&\qquad \times 
   \exp\Bigl[-\epsilon\,
   \frac{s\sqrt{s}}{\tan\sqrt{s}} +  f\,s\Bigr],
\end{eqnarray}
which simplifies drastically for the special case of fixed boundary conditions
with $\phi(0) = \phi(L) = 0$,
\begin{equation}
\label{3dp35}
   p_{0}(f) = \frac{1}{2\pi i} \int_{-i\infty}^{+i\infty}\!\!\!\!\!\!
    ds\, \Bigl(\frac{\sqrt{s}}{\sin\sqrt{s}}\Bigr)^{1/2} \!\!
   \exp\bigl(f\,s\bigr).
\end{equation}
The above result already expresses an important property of the distribution
function $p_{0}(f)$: for $f > 0$ the expression under the integral is
analytic and quickly goes to zero at $s \to -\infty$, hence the contour of
integration in the complex plane can be safely shifted to $-\infty$. This
implies that the function $p_{0}(f)$ must be equal to zero for $f > 0$
and the relaxational free energy of the directed polymer with zero boundary
conditions is bounded from above, $F < 0$.  This constraint then is easily
understood, as the presence of a random force can only {\it reduce} the
relaxational free energy of the directed polymer.
\begin{figure}[h]
   \begin{center}
   \includegraphics[width=8.0cm]{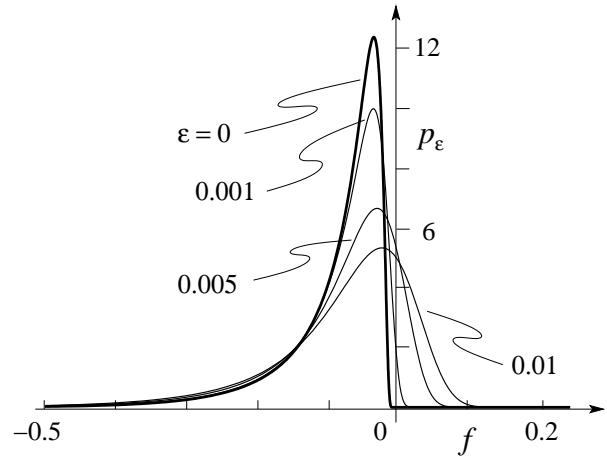}
   \caption[]{\label{fig:distrib-shift_f} Relaxational free-energy
   distribution function $p_{\epsilon}(f)$ for the randomly forced
   directed polymer for several values of the dimensionless displacement
   parameter $\epsilon = (y/\xi)^2 (L_c/L)^3/2$: $\epsilon =
   0,~0.001,~0.005,~0.01$.}
   \end{center}
\end{figure}

The evaluation of the inverse Laplace transform Eq.\ (\ref{3dp34}) is
discussed in Appendix \ref{sec:iLt} and provides the free-energy distribution
function $p_{\epsilon}(f)$ as illustrated in Fig.\
\ref{fig:distrib-shift_f} (note that all temperature dependence has vanished
since we measure our free energy with respect to the entropic contribution
$f_0^\mathrm{fs} = F_0^\mathrm{fs}/F_f = (cT/uL^2) \ln\sqrt{2\pi L T/c}$ of
the free string).  The various asymptotic forms of the distribution function
$p_{\epsilon}(f)$ in the limits $f \to \pm\infty$
and $f \to -0$ are derived via a saddle-point integration and read,
\begin{eqnarray}
\label{3dp43a}
   p_{\epsilon}(f \to -\infty) 
   &\sim& \exp\Bigl(-\pi^{2} |f|\Bigr), \\
\label{3dp43b}
   p_{\epsilon}(f \to +\infty) 
   &\sim& \exp\Bigl(-\frac{4}{27 \epsilon^{2}} [f]^{3} \Bigr), \\
\label{3dp43c}
   p_{\epsilon=0}(f \to -0) 
   &\sim& \exp\Bigl(-\frac{1}{32 |f|} \Bigr);
\end{eqnarray}
note that the shape of the left tail is determined by the singularity of
$Z(s;\epsilon)$ at $s = \pi^2$, cf.\ Eq.\ (\ref{3dp26}). The above results
agree with those obtained before in Ref.\ \onlinecite{gorokhov-blatter}.

\subsection{Distribution function: free boundary conditions}\label{sec:FrBC}

\begin{figure}[h]
   \begin{center}
   \includegraphics[width=7.0cm]{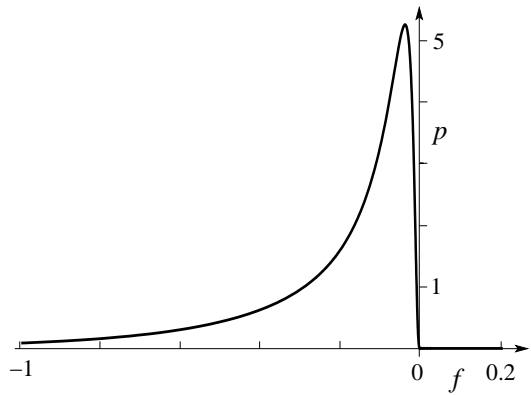}
   \caption[]{\label{fig:distrib-free_f}
   Relaxational free-energy distribution function $p(f)$ of the randomly
   forced directed polymer with free boundary conditions.}
   \end{center}
\end{figure}

The result (\ref{3dp27}) provides us with all the moments of the relaxational
free-energy distribution function, of which the first one, the average free
energy, is given by
\begin{equation}
   \label{3dp30_F}
  \overline{\langle F\rangle} = - U_c (L/L_c)^2 \,[\partial_s Z(s)]_{s=0} =
  -\frac{U_c}{4} \Bigl(\frac{L}{L_c}\Bigr)^2.
\end{equation}
In order to obtain the full distribution function, we perform the inverse
Laplace transform (${\cal P}_L(F) = p(f = F/F_f(L))/F_f(L)$)
\begin{equation}
   \label{3dp36}
   p(f) = \frac{1}{2\pi i} \int_{-i\infty}^{+i\infty}\!\!\! ds\,
   \frac{1}{\sqrt{\cos\sqrt{s}}} \, \exp( f \,s ).
\end{equation}
Given the scaling form in $f = F/F_f(L)$, the result is valid at all scales.
Again, for $f > 0$ the integrand is analytic and rapidly approaches zero as $s
\to -\infty$ and hence the function $p (f)$ must vanish identically for
$f > 0$. The functional form for $f < 0$ is found as before, see Appendix
\ref{sec:iLt}. The relaxational free-energy distribution function ${\cal
P} (f)$ assumes a universal form with no parameters; it vanishes identically
for $f > 0$ and its overall form is shown in Fig.\ \ref{fig:distrib-free_f}.

Note that the free energy of the `trivial' configuration $\phi(x) \equiv 0$ is
equal to zero and any deviation due to the action of the random force can only
reduce the energy, providing a simple explanation for the cutoff at positive
energies.  The asymptotic behavior in the limits $f \to -\infty$ and $f \to
-0$ can be easily estimated by a saddle-point calculation, 
\begin{eqnarray}
\label{3dp45a}
   p(f \to -\infty) &\sim& \exp\Bigl(-\frac{\pi^{2}}{4} |f|\Bigr),\\
\label{3dp45b}
   p(f \to -0) &\sim& \exp\Bigl(-\frac{1}{32 |f|} \Bigr).
\end{eqnarray}

\subsection{Shifted random force model}\label{sec:V}

\begin{figure}[h]
   \begin{center}
   \includegraphics[width=7.0cm]{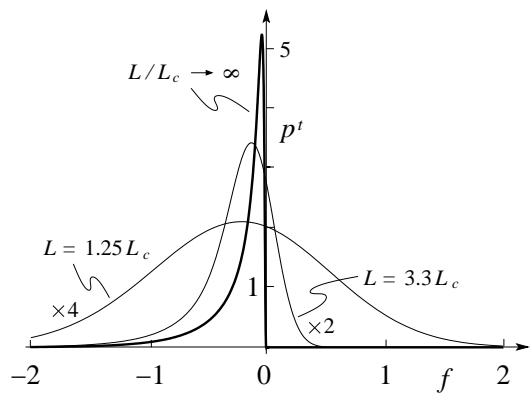}
   \caption[]{\label{fig:ran_force_shift}
   Free-energy distribution function $p^t(f)$ of the randomly
   forced directed polymer with free boundary conditions including the
   random shift $V_0(x)$. For $L \gg L_c$, the relaxational part of the
   free energy dominates the distribution; in the regime $L < L_c$, where
   the shifted random force model provides an approximation to the random
   polymer problem, the free-energy distribution function is dominated by
   the Gaussian part originating from the shift $V_0(x)$.}
   \end{center}
\end{figure}

In order to find the distribution function for the total free energy (rather
than its relaxational part), we have to account for the random shift $V_0(x)$,
cf.\ Eq.\ (\ref{1dp14_V}). Here, we concentrate on the situation with free
boundary conditions. The multiplication of Eq.\ (\ref{3dp27}) with the
Gaussian $\exp(\beta^2 n^2 U_0 L/2) = \exp[s^2(L_c/L)^3/2]$ and subsequent
Laplace transform of
\begin{equation}
   \label{Zsrf}
   Z_r(s)= \frac{1}{\sqrt{\cos\sqrt{s}}} e^{\frac{s^2}{2}\frac{L_c^3}{L^3}}
\end{equation}
generates the (rescaled) free-energy distribution function $p^t(f)$ shown in
Fig.\ \ref{fig:ran_force_shift}.

The (total) free-energy distribution function $\mathcal{P}^t$ derives from a
convolution of the distribution function ${\mathcal P}^f(F)$ of the
relaxational free energy, cf.\ Eq.\ (\ref{3dp36}) and Fig.\
\ref{fig:distrib-free_f}, and the factor $\mathcal{P}^V$ originating
from the random shift $V_{0}(x)$,
\begin{eqnarray}\label{eq:PTF}
   \mathcal{P}^t(F) = \int_{-\infty}^\infty dF' {\mathcal P}^f(F')
   \mathcal{P}^V(F-F');
\end{eqnarray}
the contribution $\mathcal{P}^V$ from the random shift assumes the simple
Gaussian form
\begin{equation}\label{eq:P}
   \mathcal{P}^V (F) = \frac{1}{\sqrt{2\pi}F_V(L)}\exp[-(F/F_V(L))^2/2]
\end{equation}
with the scaling parameter
\begin{equation}\label{eq:F_V}
   F_V(L) = U_c \sqrt{L/L_c}.
\end{equation}
For $L \gg L_c$, the result coincides with that for the relaxational energy,
cf.\ Eq.\ (\ref{3dp36}) and Fig.\ \ref{fig:distrib-free_f}, and scales as
$F/F_f(L)$. In the limit of short lengths $L \lesssim L_c$, where the model
can be used as an approximation of the random potential problem, the
distribution function is dominated by the Gaussian due to the random shift
$V_{0}(x)$. More specifically, we find the width of the distribution's body
through the calculation of the second cumulant: expanding $Z_r(s)$ for
small values of $s$ and using Eq.\ (\ref{2dp4}), we obtain
\begin{eqnarray}
  \label{Z_exp_s}
  Z_r(s) &\simeq& 1 + \frac{s}{4} + 
  \frac{s^2}{2}\Bigl[\frac{7}{48}+\Bigl(\frac{L_c}{L}\Bigr)^3\Bigr],\\
  \label{meanF}
  \overline{\langle F\rangle} &= &
  -\frac{U_c}{4} \Bigl(\frac{L}{L_c}\Bigr)^2,\\
  \label{meansF}
  \overline{F^2}-\overline{F}^2 &=& U_c^2 \frac{L}{L_c}
   \biggl[1+\frac{1}{12}\Bigl(\frac{L}{L_c}\Bigr)^3\biggr].
\end{eqnarray}
The leading term $\propto U_c/L_c^2 = u/c$ in first moment Eq.\ (\ref{meanF})
derives from the random force part of the disorder.  This is different from
the second cumulant in Eq.\ (\ref{meansF}), where the first term $\propto
U_c^2/L_c = U_0$ derives from the random shift $V_0(x)$ and dominates over the
contribution from the random force (second term $\propto (u/c)^2$) at short
lengths $L < L_c$, hence the width of ${\mathcal P}^t$ is given by $F_V$ (see
below for a discussion of the corrections $\propto (L/L_c)^3$). 

Besides the first two moments/cumulants, we can easily determine the scaling
of higher moments. Starting from the convolution Eq.\ (\ref{eq:PTF}), we note
the different scaling of the arguments in the two distribution functions,
$\propto F/L^2$ for the relaxational part $\mathcal{P}^f$ and $\propto
F/\sqrt{L}$ for the random shift part $\mathcal{P}^V (F)$. Hence, for small
distances $L$, the function $\mathcal{P}^f(F')$ peaks narrowly near zero,
while $\mathcal{P}^V(F-F')$ retains a broader shape; expanding $\mathcal{P}^V
(F-F')$ around $F$ and integrating over $F'$, we obtain the following
expansion for the total distribution function,
\begin{eqnarray}\label{eq:PTFex}
   \mathcal{P}^t (F) &\approx& \mathcal{P}^V (F)
     -{\mathcal{P}^V}'(F)\, \overline{F}^f\\
   && \nonumber \qquad\qquad\quad
     +\frac{1}{2}{\mathcal{P}^V}''(F)\, \overline{F^2}^f + \dots
\end{eqnarray}
where ${\mathcal{P}^V}'(F)$ is the derivative of $\mathcal{P}^V$ with respect
to the argument $F$ and $\overline{(\dots)}^f$ denotes averaging over the
random force part $\mathcal{P}^f$. Using the scaling $\overline{F}^f \propto u
L^2$ and $\overline{F^2}^f \propto u^2 L^4$ for the moments of the
relaxational free-energy, we can calculate the dependence of the moments
$\overline{F^{k}}$ on the length $L$. Thereby, we exploit the fact that the
leading term $\mathcal{P}^V (F)$ in the expansion (\ref{eq:PTFex}) is
symmetric in $F$, cf.\ Eq.\ (\ref{eq:P}), and hence determines the even
moments, while the next term is anti-symmetric and generates the odd moments;
finally, the third term provides the corrections to the even moments.
The combination of the scaling of $\mathcal{P}^V(F)$ (deriving from the random
shift $V_0(x)$) and of the first two moments $\overline{F}^f$ and
$\overline{F^2}^f$ (deriving from the random force $f(x)$) then generates the
following non-trivial scaling of the moments with different powers in $L$ for
the even and odd moments,
\begin{eqnarray}\label{eq:moments_e}
   \overline{F^{2k}} &\propto& L^k + \mathcal{O}(L^{k+3}), \\
   \label{eq:moments_o}
   \overline{F^{2k+1}} &\propto& L^{k+2},
\end{eqnarray}
where $\mathcal{O}(L^{k+3})$ denotes a correction term $\propto L^{k+3}$.  In
particular, $\overline{F} \propto u L^2$, $\overline{F^2} \propto U_0 L +
\mathcal{O}(u^2 L^4)$, $\overline{F^3} \propto u U_0 L^3$, $\overline{F^4}
\propto U_0^2 L^2 + \mathcal{O}(u^2 U_0 L^5)$, $\overline{F^5} \propto u
U_0^2 L^4$.  

Finally, we can estimate the tails of ${\mathcal P}^t$ from the convolution
Eq.\ (\ref{eq:PTF}) using the asymptotic behavior of ${\mathcal P}^f$ and
$\mathcal{P}^V$ and find a left tail $\mathcal{P}^t(F< -F_\mathrm{tail})
\propto \exp(F/F_f)$ and a Gaussian tail on the right, $\mathcal{P}^t(F>
F_\mathrm{tail})\propto \exp[-(F/F_V)^2/2]$, where $F_\mathrm{tail} =
F_V[1+(L_c/L)^{3/2}]$. On short scales $L < L_c$, we have $F_\mathrm{tail}
\approx F_V (L_c/L)^{3/2} > F_V$ and the random force behavior appears only
quite beyond the body.

\subsection{Joint distribution function}\label{sec:jointP}

We add a note on the joint free-energy distribution function ${\cal
P}_{L,y}(\bar F,F')$, where $\bar F = (F_{+}+F_{-})/2$ and $F' = (F_{+} -
F_{-})/2$ denote the mean free energy and the free-energy difference for two
polymer trajectories starting at the origin $\phi = 0$ at $x=0$ and ending in
the symmetric points $\phi=\pm y$ at $x=L$, $F_{\pm}\equiv F(L,\pm y;V)$.
Opposite to the $\delta$-correlated potential, cf.\ Ref.\
\onlinecite{dotsenko_08}, the present case of the random force model is less
reveiling and we keep the discussion short.

Starting with the original (random) Hamiltonian $H[\phi(x);V]$ with the random
force potential Eq.\ (\ref{1dp15}), we account for the boundary condition
$\phi(L) = y$ through the shift $\phi(x) \to (y/L) x + \phi(x)$ (the $T=0$
solution for the string ending in $\phi(L) = y$ derives from the solution
ending in $\phi(L)=0$ by adding the shift $(y/L)x$) and obtain the Hamiltonian
described by Eqs.\ (\ref{ham_bound}) and (\ref{ham_trans}).  The relaxational
free energy of the system with the boundary condition $\phi(L) = y$ separates
into the terms
\begin{equation}
\label{5dp1c}
   F[L,y;f] = \frac{c y^{2}}{2 L} + \frac{y}{L} \int_{0}^{L}\!\! dx\, x f(x)
   + F[L,0; f].
\end{equation}
The first term is the trivial part of the elastic energy, the second is a
random constant, and finally, the third is the (random) relaxational free
energy of the polymer with zero boundary conditions; its randomness is {\it
correlated} with the randomness in the second term. Then, for the free
energies $\bar F$ and $F'$ introduced above, we find that
\begin{eqnarray}
\label{5dp1d}
   F'[L,y; f] &=& \frac{y}{L} \int_{0}^{L} \!\! dx \, x f(x), \\
   \label{5dp1e}
   \bar F[L,y; f] &=& \frac{c y^{2}}{2 L} + F[L,0; f], \\ \nonumber
\end{eqnarray}
and hence $F'$ and $\bar F$ carry the information on the second and third
terms in Eq.\ (\ref{5dp1c}), respectively.  Although the joint distribution
function for the random and {\it correlated} quantities $\bar F$ and $F'$ must
be non-trivial, we can conclude that the {\it separate} statistics of $\bar F$
and $F'$ must be simple: according to Eq.\ (\ref{5dp1d}), the distribution for
$F'$ is Gaussian with zero mean and width $\overline{(F')^{2}} = y^{2} u L/3$,
while the distribution for $\bar F$ must coincide with that for the free
energy with zero boundary conditions ${\cal P}_{L,y=0}(F)$, cf.\ Eq.\
(\ref{3dp35}), shifted by the trivial elastic term $c y^{2}/2L$. Also note,
that a change in the final coordinate $y$ modifies the polymer's trajectory
over the entire length $L$ and hence the joint distribution function is not
expected to factorize, in contrast to the results found for the short-range
correlated random polymer problem \cite{dotsenko_08}.  The detailed
replica calculation, which represents a straightforward extension of the
above analysis, produces results in full agreement with these simple
arguments.

\section{Harmonic-correlator approximation}\label{sec:RPPC}

\begin{figure}[h]
   \begin{center}
   \includegraphics[width=7.0cm]{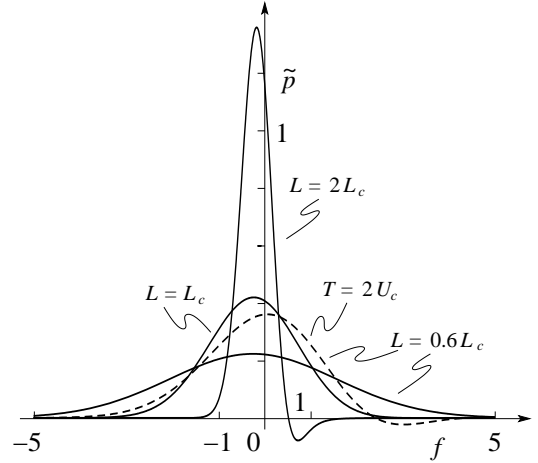}
   \caption[]{\label{fig:ran_pot_shift}
   Free-energy distribution function $\tilde{p}(f)$ for the directed polymer
   with free boundary conditions using the harmonic-correlator approximation.
   Solid curves refer to $T=0$, while the dashed curve attains to $T = 2
   U_c$.}
   \end{center}
\end{figure}

We consider the random directed polymer described by the Hamiltonian Eq.\
(\ref{1dp1}) and approximate the  interaction $-\beta^2 U$ in the replica
Hamiltonian Eq.\ (\ref{2dp6}) by the harmonic expression Eq.\ (\ref{1dp8})
to arrive at,
\begin{eqnarray}
   \label{4dp1}
   H_{n}[\{\phi_{a}\}] \!&=&\!
   \int_{0}^{L} dx \biggl\{\frac{c}{2} 
   \sum_{a=1}^{n} \bigl[\partial_x\phi_{a}(x)\bigr]^2 \\
   \nonumber
   &&\! + \frac{\beta u}{4} \sum_{a,b=1}^{n} 
   \bigl[\phi_{a}(x) -\phi_{b}(x)\bigr]^{2} \biggr\}-\frac{n^2}{2}\beta U_0 L\\
   \label{4dp2}
   &=& \nonumber
   - \frac{1}{2} \int_{0}^{L} \!\!\! dx \!\!
   \sum_{a,b=1}^{n} \! \phi_{a}(x) \, \tilde{U}_{ab} \, \phi_{b}(x)
   -\frac{n^2}{2}\beta U_0 L
\end{eqnarray}
with the matrix $\tilde{U}_{ab} = (c \partial^{2}_x - \beta n u) \delta_{ab} +
\beta u$ (note that the parabolic approximation of the correlator should be
implemented {\it after} the integration over the disorder potential).
Diagonalization produces the $(n-1)$-fold degenerate eigenvalue $\lambda_{1} =
c \, \partial^{2}_x - \beta n u$ of the harmonic oscillator problem with the
$(n-1)$ orthonormal eigenvectors $\xi^{a}_{i}$ constrained by the condition
$\sum_{a=1}^{n} \xi^{a}_{i} =0$, $i = 1, \dots, n-1$, and one non-degenerate
eigenvalue $\lambda_{2} = c \, \partial^{2}_x$ of the free problem with the
eigenvector $\xi^{a}_{n} = 1/\sqrt{n}$, $a = 1, \dots, n$.  The propagator for
the harmonic-correlator approximation then assumes the form (cf.\ (\ref{Psi}))

\begin{equation}
   \label{Psi_h}
   \Psi(\{y_a\};L) = \Bigl[\prod_{a=1}^n \Psi_{\rm h}(y_a;L)\Bigr]
   \frac{\Psi_0(\sum_b y_b/\sqrt{n};L)}{\Psi_{\rm h}
   (\sum_b y_b/\sqrt{n};L)},
\end{equation}
where $\Psi_{\rm h}$ derives from $\Psi_{\rm ih}$ by the substitution $\lambda
\to i\lambda$. For simplicity, we only consider the model with free boundary
conditions and find the shifted random force result Eq.\ (\ref{Zsrf}) replaced
by the expression ($s = L^2/\lambda^2 = n (U_c/T)^2 (L/L_c)^2$; see also
Appendix \ref{sec:nrn})
\begin{eqnarray}
   \label{Zrh}
   \tilde{Z}_r(s)
   \!&=&\!\biggl[\frac{1}{\sqrt{\cosh\sqrt{s}}}\biggr]^{(n-1)} 
   \exp\bigl[s^2(L_c/L)^3/2\bigr] \\
   \nonumber
   \!&=&\! \bigl[\cosh\sqrt{s}\bigr]^{\textstyle{\frac{1}{2}}}
   \bigl[\cosh\sqrt{s}\bigr]^{\textstyle{-\frac{s}{2}\frac{T}{U_c}
   \frac{L_c^2}{L^2}}} \,
   e^{\textstyle{\frac{s^2}{2}\frac{L_c^3}{L^3}}}.
\end{eqnarray} 
Although the inverse Laplace transform can be performed, the resulting (total)
free-energy probability distribution $\tilde{p}(f)$ develops a
negative right tail at zero and low temperatures, see Fig.\
\ref{fig:ran_pot_shift}; at large temperatures $T \gg U_c$ the right tail
exhibits pronounced oscillations. These unphysical results are due to the
departure of the approximate harmonic interaction $U_p(\phi)$ from the true
interaction $U(\phi)$, becoming relevant at large scales $L > L_c$, 
$\phi > \xi$, $F > U_c$, and the large fluctuations of the string at high 
temperatures $T \gg U_c$. Note that the inverse Laplace transform cannot 
be performed at all in case the random shift $V_0(x)$ is ignored.

The breakdown of the harmonic-correlator approximation is conveniently
observed in the second moment: expanding $\tilde{Z}_r(s)$ for small values of
$s$ we find to second order
\begin{eqnarray}
  \label{4dp13}
  \tilde{Z}_r(s) \simeq 1+\frac{s}{4}\! + \! 
  \frac{s^2}{2}\Bigl[\Bigl(\frac{L_c}{L}\Bigr)^3\!
   \Bigl(1-\frac{T}{2U_c}\frac{L}{L_c}
   \Bigr)\!-\!\frac{1}{48}\Bigr]
\end{eqnarray}
and using Eq.\ (\ref{2dp4}), we find the average free energy $\overline{F} = -
U_c(L/L_c)^2/4$ (cf.\ Eq.\ (\ref{meanF})) and the second cumulant reads
\begin{equation}
   \label{4dp14}
   \overline{F^2} - \overline{F}^2 = U_c^2 \frac{L}{L_c}
   \biggl[1-\frac{T}{2U_c}\frac{L}{L_c} 
   -\frac{1}{12}\Bigl(\frac{L}{L_c}\Bigr)^3\biggr].
\end{equation}
Comparing with the result Eq.\ (\ref{meansF}) for the shifted random force, we
note the additional dependence on temperature and the sign-change in the
correction term $\propto (L/L_c)^3$. This decrease in width is in accord with
the behavior of the free energy fluctuations in the random directed polymer
problem (\ref{1dp2}), which scales as $\delta F \propto L^{2\zeta-1}$ at large
distances; with a wandering exponent\cite{numer} $\zeta = 2/3$, we have
$\delta F \propto L^{1/3}$.  The negative correction $\propto (L/L_c)^3$ to
the linear growth in $\delta F^2$ observed in Eq.\ (\ref{4dp14}) then is
consistent with the sublinear growth $\delta F^2 \propto L^{2/3}$ of the exact
solution.

As before, we can analyze the higher moments of the distribution function and
compare the results for the harmonic-correlator approximation at $T=0$ with
those obtained for the shifted random force model. Making use of the product
form of $\tilde{Z}_r$ and expanding each factor in $s$, we find identical
leading terms for all even and odd moments (corresponding to equal expressions
for the first two terms in Eq.\ (\ref{eq:PTFex})), while the corrections to
the even moments (described by the third term in Eq.\ (\ref{eq:PTFex})) are
different.  Furthermore, we note that the even moments
$\overline{F^{2k}}\propto U_c^{2k} (L/L_c)^k [1-\mathcal{O}(L^3/L_c^3)]\propto
U_0^k L^k$ are large with a small negative correction, while the odd moments
$\overline{F^{2k+1}}\propto -U_c^{2k+1} (L/L_c)^{k+2}
[1+\mathcal{O}(L^3/L_c^3)]\propto u U_0^k L^{k+2}$ are small, their ratios
being $(\overline{F^{2k+1}})^2/\overline{F^{2(2k+1)}} \propto (L/L_c)^3$. To
leading order, the free-energy distribution function for the random potential
model at small scales then is a trivial Gaussian generated by $V_0(x)$, with a
small negative shift and a small reduction in width due to the random force
term in the potential.

At $T=0$, the second cumulant turns negative for $L>2\sqrt{3/2}\, L_c$ and the
result Eq.\ (\ref{4dp14}) makes no longer any sense, hence the harmonic
approximation to a random potential problem cannot be used on scales larger
then $\xi$ (along the transverse direction) or $L_c$ (along the longitudinal
direction); at finite temperatures the regime of validity is further reduced.

Although the results for the shifted random force remain valid at any length
$L$, we emphasize that the harmonic correlator provides a better approximation
for the behavior of the short-range correlated random polymer: Both results
agree in lowest order, providing the same first moment $\bar{F}$ due to the
random force $f(x)$ and the same leading term in the second cumulant
$\overline{F^2} - \overline{F}^2$ generated by the random shift $V_0(x)$.  The
correction $\propto (L/L_c)^3$ in the second cumulant is due to the random
force $f(x)$ and contributes with the opposite sign in the shifted random
force model as compared to the harmonic correlator approximation.  While the
shifted random force result Eq.\ (\ref{meansF}) is correct (at all scales)
when dealing with a true random force model, the correction $\propto
(L/L_c)^3$ carries the wrong sign when used as an approximation to the random
potential model and it is the result of the harmonic correlator approximation
Eq.\ (\ref{4dp14}) which should be trusted.  Indeed, the harmonic correlator
preserves the translation invariance of the problem, whereas some quadratic
terms are dropped from the shifted random force model.  Expanding the
potential to to second order,
\begin{equation}
  \label{V_sec}
  V(\phi,x) = V_0(x) + f(x) \phi(x) - \frac{1}{2} g(x) \phi^2(x),
\end{equation}
we identify the terms in $\overline{V(\phi,x)V(\phi',x')}$ with the harmonic
expansion Eq.\ (\ref{1dp8}) to obtain the correlators Eqs.\ (\ref{1dp14})
and (\ref{1dp14_V}), $\overline{V_0(x) g(x')} = u\delta(x-x')$, and vanishing
mixed terms $\overline{V_0(x)f(x')} = \overline{f(x)g(x')} = 0$. A scaling
estimate of the second moment $\overline{F^2}$ using Eq.\ (\ref{V_sec}) then
provides a leading term $\propto L$ from $V_0$ and subleading terms $\propto
L^4$ from $f$ and from $g$. The contribution from the random force provides
the positive contribution $U_c (L/L_c)^4/12$ in the cumulant Eq.\
(\ref{meansF}), while the mixed terms $\overline{V_0(x) g(x')}$ contribute
with a negative weight, generating the negative correction $\propto L^4$ in
Eq.\ (\ref{4dp14}). Note that higher order terms do not change this result
but contribute to the next order term $\propto L^7$.

Given that the harmonic correlator provides the better approximation to the
random polymer problem at short scales, one may wonder why we end up with
unphysical results (negative distribution function, negative second moment) at
larger scales. Also, different types of correlators, e.g., power-law type,
have been studied in the past, cf.\ Refs.\ \onlinecite{hh_zhang_95} and
\onlinecite{parisi_90b}, and one would like to know, what properties of a
disorder correlator guarantee consistent results; this question is addressed
in the following section.

\subsection{Correlators with non-positive spectrum}\label{sec:NSC}

It is important to identify problematic correlators right from the beginning;
indeed, the proper definition of the disorder potential is subject to
important constraints \cite{denis_99} regarding its shape $U(\phi)$ and
failure to respect these constraints may lead to unphysical results.  Consider
a random potential $V(\phi)$ and its Fourier re\-presentation $V(q) = \int
d\phi V(\phi) \exp(-iq\phi)$; then the Gaussian distribution function of the
random function $V(q)$ has the form
\begin{equation}
   \label{1dp9}
   {\cal P}[V(q)] = P_0 \exp\Bigl(-\int \frac{dq}{2\pi}\,
   \frac{|V(q)|^2}{2 G(q)}\Bigr);
\end {equation}
the width $G(q)$ has to be positive and relates to the correlation function
$U(\phi)$ via
\begin{equation}
   \label{1dp10}
   U(\phi) = \int \frac{dq}{2\pi} \, G(q) \,\exp(i q\phi).
\end {equation}
Expanding both sides in powers of $\phi$,
\begin{eqnarray}
   \label{1dp11}
   &&U(0) + \sum_{k=1}^{\infty} \frac{U^{(2k)}(0)}{(2k)!}\,\phi^{2k} \\
   \nonumber
   &&\quad = \int\! \frac{dq}{2\pi} \, G(q)
   + \sum_{k=1}^{\infty} \frac{(-1)^{k}}{(2k)!}
   \Bigl(\int\! \frac{dq}{2\pi} G(q)\, q^{2k}\Bigr)\phi^{2k},
\end{eqnarray}
and comparing coefficients, we find that the $2k$-th derivative of $U(\phi)$
in the origin relates to the integral $\int dq \, G(q) \, q^{2k}$, which is a
positive quantity. Hence, we have to be careful in our choice of the
correlator $U(\phi)$. E.g., truncating the expansion of $U(\phi)$ beyond some
$k^*$, such that $U^{(2k)}(0) = 0$ for $k \geq k^*$, we impose the condition
\begin{equation}
   \label{1dp12}
   \int dq \, G(q) q^{2k} = 0 \textrm{  for  } k\geq k^*,
\end{equation}
which cannot be satisfied for a positively defined $G(q)$.  

Obviously then, choosing a parabolic correlator $U_{p}(\phi)$ as in Eq.\
(\ref{1dp8}) is in severe conflict with the constraint Eq.\ (\ref{1dp12}). The
averaging over the disorder potential $V(\phi,x)$ is undefined for those modes
(in Fourier space) where $G(q)$ is negative. Hence, going over from the
disordered directed polymer (a statistical mechanics problem) to the quantum
boson problem is an ill-defined step and the results cannot be trusted any
longer.  On the other hand, performing the integration over the random
potential $V(\phi,x)$ with a well defined, i.e., spectral, correlator
$U(\phi)$ and expanding the resulting interaction
$-\beta^2 U(\phi)$ between bosons is perfectly admissible and leads to an {\it
identical result}; in this case, however, we know that the quantum boson
problem {\it does not} describe the random polymer problem on scales where the
approximate quadratic correlator deviates strongly from the original
correlator.  Nevertheless, in the end we have to appreciate, that the
harmonic-correlator approximation (\ref{1dp8}), although breaking down at
lengths beyond $L_c$, does produce more accurate approximate results for the
short-range correlated random potential problem (\ref{1dp2}) than the shifted
random force model (\ref{Vfphi}), although the latter remains formally valid
at all length scales $L$. The (shifted) random force model then should be used
whenever the disorder landscape is given by a force field as defined by Eqs.\
(\ref{Vfphi}), (\ref{1dp14}), and (\ref{1dp14_V}), but not as an
approximation to a random potential problem.

\subsection{Displacement correlator}\label{sec:displacement}

Another quantity of interest in the random polymer probem is the displacement
correlator $\overline{\langle\phi^{2}\rangle}(L) \equiv \overline{\langle
[\phi(L)-\phi(0)]^2 \rangle}$, with $\langle \dots \rangle$ and
$\overline{(\dots)}$ denoting thermal and disorder averages, respectively.
Choosing free boundary conditions with $\phi(0)=0$ and an arbitrary position
$\phi(L) = y$ for the end-point, the averages $\overline{\langle y^2 \rangle}$
and $\overline{\langle y \rangle^2}$ are easily calculated within replica
theory \cite{parisi_90b}.  Defining
\begin{eqnarray}
   \label{pa_pb}
   \overline{\langle y_a y_b \rangle} =
   \biggl[\, \prod_{c=1}^{n} \int dy_c \biggr] y_a y_b \Psi(\{y_c\};L),
\end{eqnarray}
we obtain the two types of averages
\begin{eqnarray}
   \label{y_corr}
   \overline{\langle y^2 \rangle} &=& \lim_{n\to 0}
   \overline{\langle y_a y_b \rangle}|_{a=b}, \\
   \nonumber
   \overline{\langle y \rangle^2} &=& \lim_{n\to 0}
   \overline{\langle y_a y_b \rangle}|_{a\neq b} .
\end{eqnarray}
The Hamiltonians for the shifted random force model and the
harmonic-correlator approximation differ only by the term $(\beta u n/2)
\sum_{a=1}^n \phi_a^2$, which vanishes in the limit $n \to 0$, hence both
schemes produce identical results for the displacement correlators in Eq.\
(\ref{y_corr}). We then concentrate on the random force case and calculate the
expression
\begin{eqnarray}
   \label{papb_cal}
   \overline{\langle y_a y_b \rangle} = C
   \biggl[\, \prod_{c=1}^{n} \int dy_c \biggr] y_a y_b 
   \exp\bigl[-\frac{1}{2}\sum_{c,d} K_{cd} \, y_c y_d\bigr]
\end{eqnarray}
with $K_{cd} = A\delta_{cd} + B$ and
\begin{eqnarray}
   \nonumber
   A &=& \frac{\beta c}{L},  \qquad
   B = \frac{\beta c}{nL} \Bigl[\frac{\sqrt{s}}{\tan\sqrt{s}}-1\Bigr],
   \\ \nonumber
   C &=& \Bigl(\frac{\beta c}{2\pi L}\Bigr)^{n/2}
       \Bigl(\frac{\sqrt{s}}{\sin\sqrt{s}}\Bigr)^{1/2}.
\end{eqnarray}
In the calculation of $\overline{\langle y_a y_b \rangle}|_{a\neq b}$, we
combine all diagonal terms into a sum $(D/2)\sum_c y_c^2$ with $D = A+B$,
leaving the non-diagonal in the form $(B/2)\sum_{c\neq d} y_c y_d$; the
non-diagonal average then follows from the derivative 
\begin{equation*}
   \overline{\langle y_a y_b \rangle}|_{a\neq b}
   = -\frac{2}{n(n-1)} \frac{\partial}{\partial B} 
   \biggl[\, \prod_{c=1}^{n} \int \! dy_c \biggr] \Psi(\{y_c\};L)\big|_D,
\end{equation*}
while the diagonal average is given by the derivative
\begin{eqnarray}
   \label{diag}
   \overline{\langle y_a^2 \rangle}
   = \frac{2}{n} \frac{\partial}{\partial A}
   \biggl[\, \prod_{c=1}^{n} \int dy_c \biggr] \Psi(\{y_c\};L)\big|_B.
\end{eqnarray}
The final results assume the form
\begin{eqnarray}
   \label{y_corr_res}
   \overline{\langle y^2 \rangle} &=& \lim_{n\to 0} 
   \Bigl(\frac{1}{A}-\frac{B}{A^2}\Bigr) = 
   \xi^2 \frac{T}{U_c}\frac{L}{L_c}+\frac{\xi^2}{3} 
      \Bigl(\frac{L}{L_c}\Bigr)^3,
   \\
   \nonumber
   \overline{\langle y \rangle^2} &=& -\lim_{n\to 0}
   \frac{B}{(D-B)^2} = -\lim_{n\to 0} \frac{B}{A^2}
   = \frac{\xi^2}{3} \Bigl(\frac{L}{L_c}\Bigr)^3;
\end{eqnarray}
the relation $\overline{\langle y^2 \rangle}-\overline{\langle y \rangle^2} =
\langle y^2 \rangle|_{V=0} = TL/c$ (here, $\langle y^2 \rangle|_{V=0}$ denotes
the thermal average in the absence of any disorder, $V=0$), is a constraint
holding true for any disorder potential uncorrelated in $x$, cf.\ Refs.\
\onlinecite{Schulz_88} and \onlinecite{Korshunov_01}.

\section{Summary and Conclusions}\label{sec:D}

The (shifted) randomly forced polymer model and the random disordered polymer
described through a harmonic-correlator approximation define quadratic problems
and hence admit exact solutions.  For the random force models, different
approaches can be taken, either a direct integration of the path-integrals
within a Fourier representation or using the (real space) replica technique;
in retrospect, the preferred method is a matter of taste.  We have determined
the free-energy distribution functions ${\cal P}_{L,y}(F)$ and ${\cal
P}_{L}(F)$ for fixed and free boundary conditions.  This calculation
necessitates the determination of all powers $Z_r(n;L,y) = \overline{Z^n
[L,y;V]}$ (rather than the usual $n\to 0$ limit) and subsequent inverse
Laplace transformation of the analytically continued replica partition
function $Z_r(n \in \mathbb{C};L,y)$.  The displacement correlators
$\overline{\langle y^2 \rangle}$ and $\overline{\langle y \rangle^2}$ have
been found as well. The simplicity of the quadratic models allows to carry
through the entire program and thus serves to study not only the physical
properties of the problem but its methodological aspects as well.

Regarding the shape of the distribution functions for the random force model,
a number of interesting features has been obtained: for the free boundary, the
probability to find a positive free energy $F$ vanishes exactly, with an
essential singularity appearing in ${\cal P}_{L}(F) \propto \exp(-uL^2/32c
|F|)$ as $F$ approaches zero from the left, cf.\ Eq.\ \ref{3dp45b}.  For fixed
boundary conditions, a similar result has been found for ${\cal P}_{L,y=0}
(F)$, see also Ref.\ \onlinecite{gorokhov-blatter}.  Furthermore, the left and
right tails provide a consistent scaling $F \propto L^2$ and $y \propto
L^{3/2}$, ${\cal P}_{L,y}(F\to -\infty) \propto \exp(-\pi^2 c |F|/u L^2)$ and
${\cal P}_{L,y}(F\to\infty) \propto \exp(-(16/27)|F|^3/u c y^4)$, cf.\ Eqs.\
(\ref{3dp43a}) and (\ref{3dp43b}).

When interested in the short distance behavior of the random directed polymer
Eq.\ (\ref{1dp2}), two types of approximations offer a drastic simplification
of the problem: these are the expansion of the random potential $V(\phi,x)$
according to Eq.\ (\ref{Vfphi}) (generating the shifted random force problem)
or the expansion of the correlator Eq.\ (\ref{1dp8}) (leading to the
harmonic-correlator approximation). While both approximations generate the
same results for the even and odd moments to leading order, the next to
leading order terms turn out different. In this situation, the results of the
harmonic-correlator approximation have to be trusted, as it consistently
accounts for the relevant terms preserving the translation invariance of the
problem.  Collecting all results, we find that the free-energy distribution
function for the random potential model at ($T=0$ and) small scales is a
trivial Gaussian of width $U_c \sqrt{L/L_c}$ generated by $V_0(x)$, with a
small negative shift $-U_c (L/2L_c)^2$ and a small reduction $-(U_c/24)
(L/L_c)^{7/2}$ in width due to the random force term in the potential.

Finally, we mention a few useful insights regarding the replica technique
which derive from our analysis above. The replica technique provides a link
between two seemingly unrelated problems, the classical statistical theory of
disordered polymers and the quantum many-body theory of attractive bosons.
Several stumbling blocks can be eliminated by properly appreciating the
subtleties in this mapping.  As is well known, after the mapping from polymers
to bosons the disorder correlator assumes the role of the interaction
potential. While many shapes for the interaction potential may produce
meaningful results for the quantum boson problem, only a restricted set of
them (those describing a correlator with positive spectrum) relate to a
meaningful random polymer problem. Hence, the original choice of physical
correlators and any modification thereof during the calculation should be done
with great care; in particular, a simple power-law form \cite{hh_zhang_95}
might not work.  E.g., there is nothing wrong in studying quantum bosons with
a simple harmonic interaction $U(\phi) = -U_0 + u \phi^2/2$ and the results
obtained for the quantum propagator are perfectly acceptable for any constant
shift $U_0$.  However, interpreting the result for the propagator in terms of
a replica partition function and transforming back (via the inverse Laplace
transformation) to random polymers, the resulting distribution function
becomes unphysical when setting $U_0 = 0$; dropping a shift $U_0$ in the
potential for the bosons is a trivial shift in energy, while ignoring the same
shift in the correlator produces unphysical results for the polymer problem
after Laplace transformation.

\acknowledgments

We thank Sergey Korshunov for numerous enlightening discussions and
acknowledge financial support from the Pauli Center at the ETH Zurich and the
Swiss National Foundation.

\appendix

\section{Negative replica number}\label{sec:nrn}

We determine the replica partition function $Z_r(s)$, Eq.\ (\ref{3dp30}), for
the polymer with free boundary conditions via direct solution of the
Schr\"odinger equation Eq.\ (\ref{3dp18}) for negative $n$. We confirm, that
the result analytically continued from positive $n$ agrees with the one
obtained for negative $n$.  The wave function $\Psi(q,x)$ satisfies the
Schr\"odinger equation (cf.\ Eq.\ (\ref{3dp18}))
\begin{equation}
   \label{Adp1}
   \partial_x \Psi(q; x) = \frac{1}{2}\Bigl[\frac{1}{\beta c} \partial_{q}^2 
              - \frac{\beta c}{\lambda^2}\, q^2\Bigr] \, \Psi(q;x),
\end{equation}
with $\lambda^2 = c/\beta |n| u$.  We are seaking the solution 
\begin{equation}
   \label{Adp4}
   \Psi(q; x) = \sum_{k=0}^{\infty} A_{k} \, \exp(-E_{k} x) \, \Psi_{k}(q),
\end{equation}
satisfying the initial condition $\Psi(q;x=0) =\delta(q)$; the energies and
corresponding orthonormal eigenfunctions $E_{k}$ and $\Psi_{k}(q)$ 
satisfy the stationary equation,
\begin{eqnarray}
   \label{Adp5}
   E_{k} \Psi_{k}(q) = -\frac{1}{2}\Bigl[\frac{1}{\beta c} \partial_{q}^2 
                     + \frac{\beta c}{\lambda^2}\, q^{2}\Bigr] \, \Psi_{k}(q).
   \\ \nonumber 
\end{eqnarray}
The coefficients $A_{k}$ in Eq.\ (\ref{Adp4}) derive from the initial
condition
\begin{equation*}
   A_{k} =  \int_{-\infty}^{+\infty} \!\! dq \, \Psi_{k}^{*}(q) \Psi(q; x=0) 
   = \Psi_{k}^{*}(0).
\end{equation*}
The spectrum of the harmonic problem is given by $E_{k} = (k+1/2)/\lambda$ and
the corresponding eigenfunctions are (see, e.g.. Ref.\
\onlinecite{Landau-Lifshitz})
\begin{equation*}
   \Psi_{k}(q) = \Bigl(\frac{\beta c/\lambda}{\sqrt{\pi} 2^{k} k!}\Bigr)^{1/2}
   \exp[-(\beta c/2\lambda)q^2] 
   H_{k}[\sqrt{\beta c/\lambda} q],
\end{equation*}
where $H_{k}(x)$ are the Hermite polynomials $H_{k}(x) = (-1)^k \exp(x^2) \,
\partial^k_x [ \exp(-x^2)]$.  Substituting $A_k$ and $\Psi_k$ into Eq.\
(\ref{Adp4}) and taking into account that $H_{2l+1}(0) = 0$, we obtain the
wave function
\begin{eqnarray}
   \label{Adp11}
   &&\Psi(q; x) = \sqrt{\frac{\beta c}{\pi \lambda}}
   \sum_{l=0}^{\infty}\frac{1}{ 2^{2l}(2l)!}
   \, \exp(-E_{2l} x)\, \\
   \nonumber
   &&\qquad\times\exp[-(\beta c/\lambda)\, q^2] \,
   H_{2l}[\sqrt{\beta c/\lambda} \, q] \, H_{2l} (0).
\end{eqnarray}
With the spectrum $E_k(\lambda)$ and the normalization $H_{2l}(0) =
(-1)^{l} (2l)!/2^l l!$, we obtain the replica partition function for free
boundary conditions (cf.\ Eq.\ (\ref{3dp27}))
\begin{eqnarray}
   \label{Adp12}
   Z_r(n;L) &\equiv& Z(s) = \int_{-\infty}^{+\infty} \!\! dq \, 
   \Psi(q; x=L) \\
   \nonumber
   &=& \frac{\exp(-\sqrt{s}/2)}{\sqrt{\pi}} \, 
   \sum_{l=0}^{\infty} \frac{(-1)^{l}}{2^{2l} \, l!} \, 
   \exp(- 2 \sqrt{s} \, l) \, C_{2l}
\end{eqnarray}
where $s = L^2/\lambda^2$ and 
\begin{equation}
   \label{Adp13}
   C_{k} = \int_{-\infty}^{+\infty} \!\! dx \, 
   \exp(-x^{2}/2) \; H_{k}(x).
\end{equation}
Using the recurrence relation $H_{k+1}(x)=2x \, H_{k}(x)-2k
H_{k-1}(x)$, we find that $C_{k+2} = 2(k+1) C_{k}$ and with $C_{0} =
\sqrt{2\pi}$, we obtain the coefficients $C_{k} = \sqrt{2\pi} (2l)!/ l!$.
Substitution into Eq.\ (\ref{Adp12}) provides the replica
partition function in the form 
\begin{equation}
   \label{Adp17}
   Z(s) = \sqrt{2} \, \exp(-\sqrt{s}/2) \,  R[\eta(s)]
\end{equation}
with the function $R(\eta)$ defined by the series
\begin{equation}
   \label{Adp18}
   R(\eta) = \sum_{l=0}^{\infty} \frac{(2l)!}{(l!)^{2}} \eta^{l}
\end{equation}
and we have introduced the shorthand $\eta(s) = -\exp(-2\sqrt{s})/4$.

In order to find the explicit form of the function $R(\eta)$, we implement
the shift $l \to l+1$ in the sum (\ref{Adp18}) and obtain,
\begin{eqnarray}
   \nonumber
   R(\eta) &=& 1 + \sum_{l=1}^{\infty} \frac{(2l)!}{(l!)^{2}} \eta^{l} 
            = 1 +  \sum_{l=0}^{\infty} \frac{(2l+2)!}{(l!)^{2}}\eta^{l+1} 
   \\ \nonumber
   &=& 1 + 4\eta \sum_{l=0}^{\infty} \frac{(2l)!}{(l!)^{2}} \eta^{l} 
       - 2 \sum_{l=0}^{\infty} \frac{(2l)!}{(l+1)(l!)^{2}} \eta^{l+1} 
   \\ \label{Adp20}
   &=& 1 + 4\eta R(\eta) - 2 S(\eta),\\
   \label{Adp21}
   S(\eta) &=& \sum_{l=0}^{\infty} \frac{(2l)!}{(l+1)(l!)^{2}} \eta^{l+1}.
\end{eqnarray}
With $R$ the derivative of $S$, $R(\eta) = \partial_\eta S(\eta)$, we obtain
the differential equation $\partial_\eta S(\eta) = 1 + 4 \eta \partial_\eta
S(\eta)-2 S(\eta)$ and the initial condition $S(0) = 0$ determines the
solution $S(\eta) = (1 - \sqrt{1-4 \eta})/2$, from which $R(\eta) = 1/\sqrt{1
- 4 \eta}$ follows via integration.  Substitution into Eq.\ (\ref{Adp17})
produces the final result $Z(s) = {1}/{\sqrt{\cosh\sqrt{s}}}$, in agreement
with Eq.\ (\ref{3dp30}).

\section{Inverse Laplace transformations}\label{sec:iLt}

The inverse Laplace transforms Eqs.\ (\ref{3dp34}) and (\ref{3dp36}) are
reduced to the following expressions: Using the transformation $s = \rho
\exp(\pm i\pi/2)$ in Eq.\ (\ref{3dp34}), we analytically continue the
expression for the distribution function $p_{\epsilon}(f)$ to the
imaginary axis,
\begin{eqnarray} \label{3dp37}
   p_{\epsilon}(f) 
   &=& \frac{1}{\pi} \mathrm{Re} \int_{0}^{\infty} \!\!\!\! d\rho \,
   \biggl(\frac{\sqrt{\rho}\exp(i\pi/4)}
   {\sin[(1+i)\sqrt{{\rho}/{2}}]}\biggr)^{1/2}
   \\ \nonumber &&
   \times \exp\biggl[\epsilon \frac{(1-i) \rho\sqrt{\rho/2}}{
   \tan[(1+i)\sqrt{{\rho}/{2}}]} +if\rho\biggr].
\end{eqnarray}
A change of the integration variable $\rho = 2 t^{2}$ provides, after some
algebra, the final expression
\begin{eqnarray}
\label{3dp38}
   p_{\epsilon}(f) &=& \frac{2^{5/2}}{\pi} 
   \int_{0}^{\infty}\!\!\! dt \, t^{3/2} \exp[-\epsilon\omega_{-}(t)] \\
   \nonumber
   && \times \frac{\cos[\gamma(t)/2+2 t^2 f+{\pi}/{8}
   -\epsilon \omega_{+}(t)]}{\sqrt{\Phi(t)}}.
\end{eqnarray}
The functions $\Phi(t)$, $\omega_{\pm}(t)$, and $\gamma(t)$ are defined as,
\begin{eqnarray}
   \nonumber
   \Phi(t) &=& \sqrt{[\sin(t) \cosh(t)]^{2} + [\cos(t) \sinh(t)]^{2}}, \\
   \nonumber
   \omega_{\pm}(t) &=& t^3 [\sinh(2t) \pm \sin(2t)]/\Phi^{2}(t), \\
   \nonumber
   \sin(\gamma(t)) &=& -\cos(t) \sinh(t)/\Phi(t), \\
   \nonumber
   \cos(\gamma(t)) &=&  \sin(t) \cosh(t)/\Phi(t).
\end{eqnarray}
Similarly, substituting $s = 2 t^{2} \exp(\pm i \pi/2)$ in Eq.\ (\ref{3dp36}),
one obtains
\begin{eqnarray}
\label{3dp44}
   p(f) &=& \frac{4}{\pi} \int_{0}^{\infty} dt \, t
   \,\frac{\cos[\zeta(t)/2 + 2f t^{2}]}{\sqrt{\Psi(t)}},
\end{eqnarray}
with the functions $\Psi(t)$ and $\zeta(t)$ defined by 
\begin{eqnarray}
   \nonumber
   \Psi(t) &=& \sqrt{[\cos(t)\cosh(t)]^{2} + [\sin(t)\sinh(t)]^{2}}, \\
   \nonumber
   \sin(\zeta(t)) &=& \sin(t) \sinh(t)/\Psi(t),
   \\ \nonumber
   \cos(\zeta(t)) &=& \cos(t) \cosh(t)/\Psi(t).
\end{eqnarray}
The remaining integrals in Eqs.\ (\ref{3dp38}) and (\ref{3dp44}) have to be
done numerically.

\end{document}